\documentstyle[12pt,epsf]{article}
\newcommand{\be}{\begin{equation}}
\newcommand{\ee}{\end{equation}}
\newcommand{\bea}{\begin{eqnarray}}
\newcommand{\eea}{\end{eqnarray}}
\newcommand{\vap}{\varepsilon}
\newcommand{\pary}{\partial_{y}}
\newcommand{\parp}{\partial}

\title{\hfill $\mbox{\small{
$\stackrel{\rm\textstyle SU{-}4240{-}680\quad}
{\rm \textstyle \quad}
$}}$ \\[1truecm]
The tubular phase of self-avoiding anisotropic crystalline membranes}
\author{\\\\ Mark Bowick\thanks{\tt bowick@physics.syr.edu} \, 
and Alex Travesset\thanks{\tt alex@suhep.phy.syr.edu}
\\\\ Physics Department, Syracuse University,\\
Syracuse, NY 13244-1130, USA \\ }

\date{}
\begin{document}

\begin{titlepage}
\maketitle
\begin{abstract}
We analyze the tubular phase of self-avoiding anisotropic crystalline 
membranes. A careful analysis using renormalization group arguments
together with symmetry requirements motivates the simplest form of the
large-distance free energy describing fluctuations of tubular
configurations. The non-self-avoiding limit of the model is shown to
be exactly solvable. For the full self-avoiding model we compute the
critical exponents using an $\vap$-expansion about the upper critical
embedding dimension for general internal dimension $D$ and embedding dimension $d$.
We then exhibit various methods for reliably extrapolating to the
physical point $(D=2,d=3)$. Our most accurate 
estimates are $\nu=0.62$ for the Flory exponent and $\zeta=0.80$
for the roughness exponent.    
\end{abstract}
\end{titlepage}

\section{Introduction}\label{SECT__Introduction}

The statistical mechanics of isotropic crystalline membranes has been
the subject of much work in the last ten years \cite{Review1, Review2}. 
In the absence of {\em self-avoidance} there is a finite temperature 
{\em crumpling} transition from a low-temperature flat (orientationally-ordered) phase
to a high-temperature crumpled phase. The novel flat phase of phantom
crystalline membranes is by now quite well understood, both
qualitatively and quantitatively. The effect of self-avoidance on the phase
diagram presents a much greater analytical and numerical challenge.
While there is still some controversy, the bulk of evidence at present 
indicates that the crumpled phase disappears. It is possible, however, that this is
the result of bending rigidity induced by next-to-nearest-neighbor 
excluded volume interactions.

Rather surprisingly, it has been shown \cite{RT1} that anisotropy has a 
remarkable effect on the global phase diagram of this class of
membranes. For phantom membranes the flat and crumpled phases are
isomorphic to those of the isotropic system (anisotropy is irrelevant 
in these phases) but there are intermediate tubular phases
in which the membrane is ordered in one extended direction (y) and crumpled
in the remaining transverse directions ($\perp$).
Since self-avoidance is less constraining for configurations that are
crumpled in one direction only, it is very likely that the tubular
phase will survive in the more physical self-avoiding case, in
contrast to the situation for isotropic membranes.
Besides their intrinsic novelty, the study of membranes of this class
may have important experimental and practical applications.
First of all polymerized membranes with in-plane tilt order would have
intrinsic anisotropy. In addition, polymerization in the presence of an 
applied electric field should produce anisotropic membranes \cite{Bensimon}. 

\begin{figure}[htb]
\epsfxsize=5in \centerline{\epsfbox{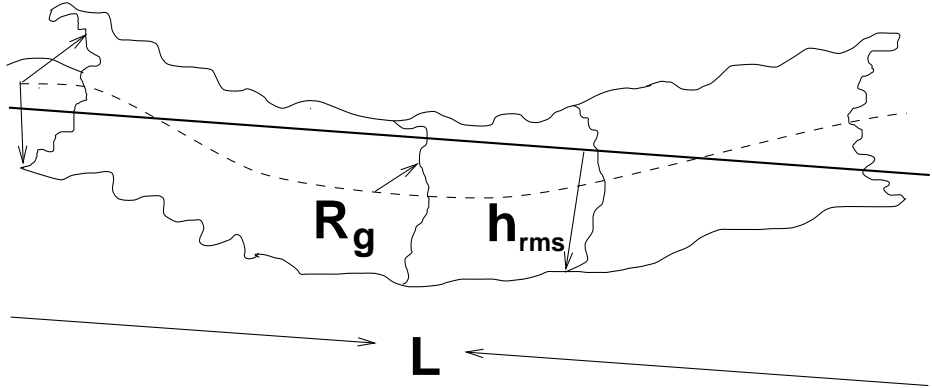}}
\caption{A schematic illustration of a tubular configuration
indicating the radius of gyration $R_g$ and the height fluctuations $h_{rms}$.} 
\label{fig__tubdef}
\end{figure}

The key critical exponents characterizing the tubular phase are 
the size (or Flory) exponent $\nu$, giving the scaling of the tubular
diameter $R_g$ with the extended $L_y$ and transverse $L_{\perp}$
sizes of the membrane, and the roughness exponent $\zeta$ associated with the growth of height
fluctuations $h_{rms}$ (see Fig.~\ref{fig__tubdef}):
\bea
\label{nuzeta}
R_g(L_{\perp},L_y) & \propto & L_{\perp}^{\nu} S_R(L_y/L_{\perp}^z)\\
\nonumber
h_{rms}(L_{\perp},L_y) & \propto & L_y^{\zeta} S_h(L_y/L_{\perp}^z)
\eea 
Here $S_R$ and $S_h$ are scaling functions \cite{RT1,RT2}
and $z=\frac{\nu}{\zeta}$ is the anisotropy exponent.
In the phantom tubular phase (PTP) $\nu$ and $\zeta$ were computed in
\cite{RT1}, together with a self-consistent determination of the
anomalous elasticity. The existence of the tubular phase has also been
confirmed by numerical simulations \cite{BFT} and the critical
exponents measured are in excellent agreement with the theoretical predictions.
In this paper we show that a careful analysis of the relevant operators in the 
free energy allows an {\em a priori} exact calculation of the anomalous
elasticity as well as the above critical exponents.

\begin{figure}[htb]
\epsfxsize=5in \centerline{\epsfbox{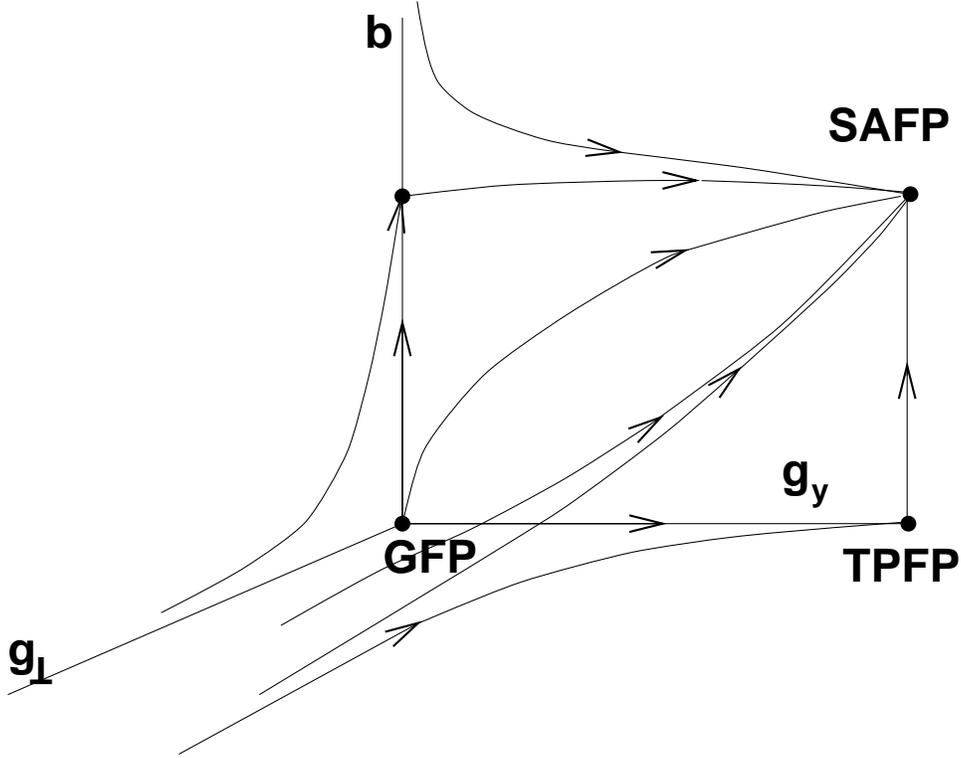}}
\caption{The phase diagram for self-avoiding anisotropic membranes
with the Gaussian fixed point (GFP), the tubular phase fixed point
(TPFP) and the self-avoidance fixed point (SAFP).} 
\label{fig__BG}
\end{figure}

\begin{figure}[htb]
\epsfxsize=5in \centerline{\epsfbox{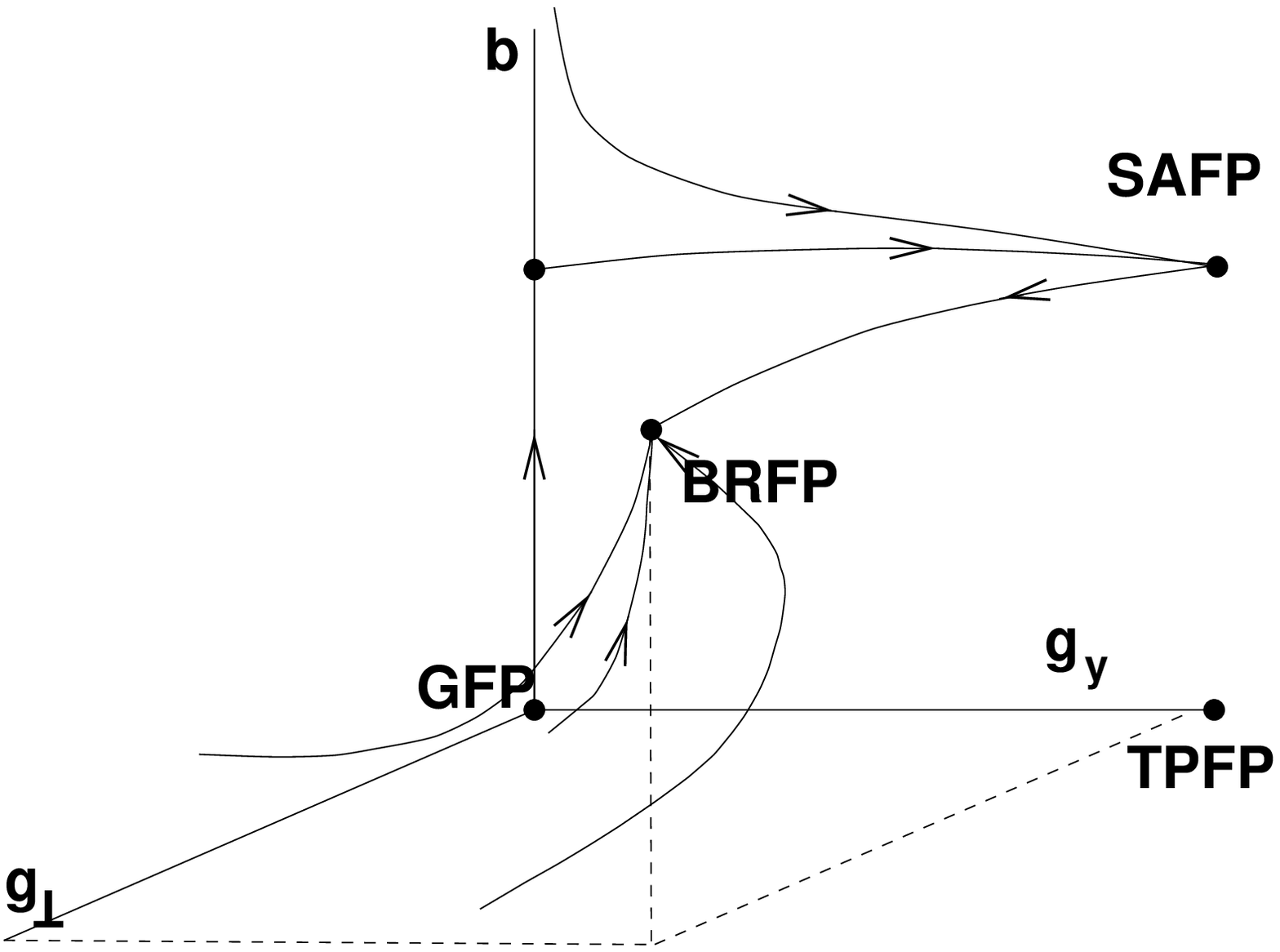}}
\caption{The phase diagram for self-avoiding anisotropic membranes
with the Gaussian fixed point (GFP), the tubular phase fixed point
(TPFP), the self-avoidance fixed point (SAFP) and the bending rigidity
fixed point (BRFP).} 
\label{fig__RT}
\end{figure}

For self-avoiding membranes the model is much more difficult to treat 
analytically. By adapting the Edwards model
for self-avoiding membranes to the geometry of the tubular phase,
Radzihovsky and Toner \cite{RT1} obtained a model free energy to
describe this system. This was further studied by Bowick and Guitter
\cite{BG}, who utilized the multi-local-operator-product-expansion 
(MOPE)\cite{MOPE1,MOPE2} to perform an $\vap \equiv (d_c^{SA}-d)$ 
expansion about the upper critical (embedding) dimension $d_c^{SA}=11$. 
The phase diagram implied by this analysis is shown in Fig.~\ref{fig__BG}.
Note the infrared stable fixed point (SAFP) with non-vanishing
self-avoidance coupling $b$ associated with the tubular
phase. Bowick and Guitter also showed that the bending rigidity
is not renormalized and computed the critical exponents to
first order in $\vap$. They noted, however, that the extrapolation of
these predictions to the physical tubule was not very robust against  
higher-order perturbations.
   
Radzihovsky and Toner \cite{RT2} have argued that the phase diagram
described above is actually more complicated (see Fig.~\ref{fig__RT})
for embedding dimension $d$ less than a critical value $d_*$,
with $d_*>3$. They argue that the physics below $d^*$ is
controlled by a new fixed point (BRFP) which is non-perturbative in 
$\vap$. This postulated fixed point is quite distinct physically from
the SAFP. In particular the bending rigidity picks up a non-zero anomalous exponent.
Calculating critical exponents at the putative BRFP would present the formidable
challenge of a complete treatment of both self-avoidance and full
non-linear elasticity. Reasonable estimates of $\nu$ may, however, be obtained
within the Flory approximation. 

In the present paper we begin with a careful analysis of the
rotational symmetries of the tubular problem and their realization 
within a Wilsonian renormalization group approach \cite{WK}.
This constrains the possible operators that may appear in the
free energy and allows us to identify some operators as definitely 
being irrelevant with respect to a broad category of fixed points. 
As a result of our analysis we can motivate the phase diagram 
Fig.~\ref{fig__BG}, which follows from the free energy studied in \cite{BG}, 
with the incorporation of a relevant operator involving in-plane
phonon excitations. The analysis of \cite{RT2} assumes that 
non-linear elasticity terms are always irrelevant. It may therefore
break down if new terms in the free energy alter the renormalization
group flows. While this may change the character of the fixed
point above, our analysis suggests that it is imperative to understand
the SAFP in as much detail as possible. This is the focus of the present paper.
Given the model we next turn to the actual calculation of reliable 
critical exponents in the tubular phase. 
This is done by generalizing the calculation of \cite{BG} to
manifolds of arbitrary internal dimension $D$ embedded in general
dimension $d$. We analyze a class of generalized $\vap${-}expansions 
that allow us to determine an optimal path from the line $\vap=0$ to the 
physical point $(D=2,d=3)$.  
Our most accurate estimates are
\bea\label{sum_res}
\nu&=&0.62 
\nonumber \\
\zeta&=&0.80 \ .
\eea
Furthermore, we show that the critical exponents determined in this
method are extremely close to the Flory prediction, particularly for $d>3$.
This may be regarded as strengthening the predictions of the otherwise
uncontrolled Flory approximation.

The outline of our paper is as follows. The model is described in
Sec.~\ref{SECT__Model} along with an analysis of its symmetries and
their implementation in a Wilsonian renormalization group framework.
This leads to a clarification of the global phase diagram and a
proposal for the simplest free energy capturing the essential large distance
physics of the tubular phase. 
This is followed in Sec.~\ref{SECT__scalrel} by a derivation of the
scaling relations connecting the fundamental critical exponents.
The special case of the phantom tubule is treated in detail in
Sec.~\ref{SECT__phtub}. The full physical problem of the
self-avoiding tubule is tackled in Sec.~\ref{SECT__SAtub}, where 
critical exponents are computed via a generalized $\vap$-expansion.
We also compute corrections to the Flory and Gaussian variational 
approximations. A brief summary of our results is given in 
Sec.~\ref{SECT__Con}. Finally, some technical details of the 
$\vap${--}expansion are left to the Appendix Sec.~\ref{SECT__App}.
 
\section{Model}
\label{SECT__Model}

A membrane configuration may be characterized by giving the position 
$\vec r ({\bf x})$, in the $d$-dimensional embedding space, of a point in
the membrane labeled by a $D$-dimensional internal coordinate ${\bf x}$. 
A physical membrane corresponds to the case $d=3$ and $D=2$.

In \cite{RT1,RT2} the most general Landau-Ginzburg-Wilson free energy $F$ 
for this system is constructed by expanding $F$ to leading order in
powers of $\vec r(\bf {x})$ and its gradients with respect to internal
space $\bf {x}$, taking into account global translation and rotational
invariance. We will consider the case in which the membrane is
isotropic in $D-1$ membrane directions (denoted $\bf {x}_{\perp}$)
orthogonal to a distinguished direction $y$. The resultant free energy
is given by 
\bea\label{LGW}
F(\vec r({\bf x}))&=& \frac{1}{2} \int d^{D-1}{\bf x}_{\perp} dy \left[
\kappa_{\perp}(\partial_{\perp}^2 \vec r)^2 + \kappa_y (\pary^2 \vec
r)^2  \right.
\nonumber\\
&& + \kappa_{\perp y} \pary^2 \vec r \cdot \parp_{\perp}^2 \vec r +
t_{\perp}(\parp_{\alpha}^{\perp} \vec r)^2 + t_y(\pary \vec r)^2 
\nonumber\\
&& + \frac{u_{\perp \perp}}{2}(\parp_\alpha^{\perp} \vec r \cdot
\parp_{\beta}^{\perp} \vec r)^2 + \frac{u_{yy}}{2}(\pary \vec r \cdot
\pary \vec r)^2
\nonumber\\
&& + u_{\perp y} (\parp_{\alpha}^{\perp} \vec r \cdot \pary \vec r)^2
+ \frac{v_{\perp \perp}}{2}(\parp_{\alpha}^{\perp} \vec r \cdot
\parp_{\alpha}^{\perp} \vec r)^2
\nonumber\\
&& \left.  + v_{\perp y}(\parp_{\alpha}^{\perp} \vec r)^2 
(\pary \vec r)^2 \right]
\nonumber\\
&&+ \frac{b}{2} \int d^D {\bf x} \int d^D {\bf x}^\prime \delta^d 
(\vec r({\bf x}) - \vec r({\bf x}^\prime)),
\eea
where the parameters denote bending and elastic moduli.
Note the complexity of this model {--} it has eleven free parameters.
In mean field theory the non-self-avoiding limit (b=0)  
yields a phase diagram with flat and crumpled phases separated by a
tubular phase \cite{RT1}.

In this paper we will be mainly concerned with the tubular phase
(TP) beyond mean field theory. In this case we may expand $\vec r$ in
the Monge representation:
\be\label{Tub}
\vec r({\bf x})=(\zeta_{y} y + u({\bf x}), \vec h({\bf x})).
\ee
The free energy is now a function of $u$ and $\vec h$. Before 
simplifying Eq.~\ref{LGW} let us discuss the symmetries of the tubular
phase.  

Since the free energy must be invariant under global 
rotations of the tubule it is expressible in terms
of the complete set of tubular rotationally invariant operators. These
are
\bea\label{inv_op}
E(u,h)&=&\pary u+\frac{1}{2}(\pary \vec h)^2+\frac{1}{2}(\pary u)^2
\nonumber\\
F_{\alpha}(u,h)&=& \parp_{\alpha}u +\pary \vec h \parp_{\alpha} \vec h
+\pary u \parp_{\alpha} u
\nonumber\\
F_{\alpha \beta}(u,h)&=&\parp_{\alpha} u \parp_{\beta} u+
\parp_{\alpha}{\vec h} \parp_{\beta}{\vec h}
\\\nonumber
G_y(u,h)&=& (\pary^2 u)^2+(\pary^2 \vec h)^2
\nonumber\\\nonumber
G_{y \alpha \beta }(u,h)&=& (\pary^2 u)(\parp_{\alpha \beta} u)+
\pary^2 \vec h \parp_{\alpha \beta} \vec h \ .
\eea
Indeed, Eq.~\ref{LGW} becomes
\bea
\label{freeenergy}
F(u,\vec h)&=& \frac{1}{2} \int d^{D-1}{\bf x}_{\perp} dy \left[
2\zeta_{y}(t_{y}+u_{yy}\zeta_{y}^2)E(u,h)+\kappa_{y}
\zeta^4_y G_y(u,h) \right. 
\nonumber\\
&& + \kappa_{y\perp} 
G^{\alpha}_{y \alpha}(u,h)+2u_{yy}\zeta^4_{y}E^2(u,h)+(t_{\perp}+
v_{\perp y})\zeta^2_y F^{\alpha}_{\alpha}+
\zeta^4_y u_{\perp y} F_{\alpha} F^{\alpha} 
\nonumber\\
&&   \left.
+2v_{\perp y} \zeta^2_y E(u,h) F^{\alpha}_{\alpha}+
\frac{u_{\perp \perp}}{2} F^{\alpha}_{\beta} F^{\beta}_{\alpha}
+\frac{v_{\perp \perp}}{2}(F^{\alpha}_{\alpha})^2 \right]
\nonumber\\
&&+ \frac{b}{2}\int d^{D-1}{\bf x}_{\perp} dy 
d^{D-1}{\bf x}_{\perp}^\prime dy^\prime 
\delta^{d-1}(\vec h({\bf x}_{\perp},y)-\vec h({\bf x}_{\perp}^\prime,y^\prime))
\nonumber\\
&&
\times \delta(\zeta_y (y - y^\prime) + u({\bf x}_{\perp},y) - 
u({\bf x}_{\perp}^\prime,y^\prime))
\eea

Since we are interested in the critical properties of the free energy 
Eq.~\ref{freeenergy} we may simplify by dropping irrelevant terms. 
Simple power counting around the Gaussian fixed point is usually
enough to determine the relevancy of operators but in this case
the situation is more involved and requires a careful analysis of the 
symmetries of the problem, to which we turn now.

\subsection{Wilson RG in the tubular phase}\label{SUBSECT__Wilson}

We apply the Renormalization Group (RG) a la Wilson \cite{WK} 
to the free energy Eq.~\ref{freeenergy}. While this approach is usually 
more involved for extracting actual numbers than the more conventional
field theory approach \cite{AM}, it is more general and allows an easier 
analysis of the irrelevant operators, key to deciding 
which terms to retain in the free energy. The crucial point in 
Wilson RG is the RG transformation. This is a two step procedure: 
the {\em blocking} and the {\em rescaling}. 

There is considerable freedom in the choice of blocking.
We chose decimation in momentum space, where in order 
to simplify the calculations an anisotropic  spherical momentum 
regularization is assumed. The blocking just consists in integrating 
over an anisotropic shell of thickness $e^{-l} \ , \  l \in [0,+\infty)$. 
That is,
\be\label{RG_block}
e^{-F_{l}(u,\vec h)}=\int \prod_{\{ |q_{\perp}|,|q_y| \} \in {\cal B} } 
du(q_{\perp},q_y) d{\vec h}(q_{\perp},q_y)    
e^{-F(u,h)} \ .
\ee
The region ${\cal B}$ consists of three sectors 
\be\label{region_blocking}
{\cal B}=\left\{ \begin{array}{c} 1 > |q_{\perp}| > e^{-l} \ , \
            e^{-zl} > |q_y| > 0 \\ 1 > |q_{\perp}| > e^{-l} \ , \   
            1 > |q_y| >  e^{-zl} \\  
            e^{-l} < |q_{\perp}| < 0 \ , \ 1 >  |q_y| > e^{-zl}   
         \end{array} \right.
\ee
where the exponent $z$ accounts for the anisotropy of the system.
This blocking is very similar to the one used in \cite{RT2}.

The rescaling is anisotropic as well and is given by,
\be\label{RG_scal}
\begin{array}{c} q_{\perp}^\prime=e^{l}q_{\perp} \\ q_y^\prime=e^{zl}q_y 
\end{array}
\  \
\begin{array}{c} h^\prime(q^\prime)=e^{-(D-1+z+\nu)l} h(q) \\ 
u^\prime(q^\prime)= e^{-(D-1+2\nu)l} u(q)  \ , \end{array}
\ee
where $\nu$ is the other exponent that appears in the theory.

The result of performing a renormalization group transformation up
to time `l', is the Wilsonian free energy
\be\label{free_WILS}
F_l(u^\prime,\vec h^\prime) \ ,
\ee
where the $u^\prime$ and $\vec h^\prime$ fields have the same range as 
the original ones. The free energy evaluated at $l=0$ is, 
by definition, Eq.~\ref{freeenergy}.

For future reference, let us work out the simplest fixed point in 
Eq.~\ref{freeenergy}, the Gaussian fixed point. Although this fixed
point is not of direct physical interest it plays a central role in
many considerations (see Fig.~\ref{fig__BG}).
This fixed point may be studied by retaining only the quadratic
terms in the free energy Eq.~\ref{freeenergy}, and applying the 
RG transformation just defined.
We easily get (hereafter dropping all primes in the rescaling)
\bea\label{RG_quad}
F_{l}&=&\frac{1}{2}\int d^{D-1}\hat{q}_{\perp} d\hat{q}_y
\left[ (e^{(D-1-3z+2\nu) l} \kappa q^4_y+e^{(D-3+z+2\nu)l}t q^2_{\perp})
h(q)h(-q) \right.
\nonumber\\
&&+\left.
(e^{(D-1-3z+4\nu) l} g_y q^2_y+e^{(D-3-z+4\nu) l}g_{\perp} q^2_{\perp})
u(-q)u(q) \right] \ .
\eea
Imposing that the Gaussian fixed point is given by the terms 
involving $\vec h$, 
the exponents $z$ and $\nu$ are readily computed to be
\be\label{Gexp}
z=\frac{1}{2} \ , \  2 \nu = \frac{5}{2}-D \ 
\ee
and the exponents for the operators associated with the couplings 
are uniquely determined. The Gaussian fixed point is thus
$g_y=g_{\perp}=0$. 
The coupling $g_{\perp}$ defines an irrelevant direction for $D>3/2$, 
with exponent $\frac{3}{2}-D$, while
$g_y$ defines a relevant direction for $D<5/2$, with exponent 
$\frac{5}{2}-D$. The Gaussian fixed point is therefore infrared unstable.

\subsection{The rotations of the tubule}\label{SUBSECT__rotations}

For the general free energy of Eq.~\ref{freeenergy} the rotations of the
tubule are implemented by 
\be\label{exact_rot}
\begin{array}{l l} u \rightarrow  & u \cos \theta + \sin \theta h +
(\cos \theta -1)y \\
h \rightarrow   & h \cos \theta - \sin \theta u -\sin \theta y
\end{array} \ ,
\ee
where we have simplified by rotating just one component of $\vec h$.
The symmetry transformation above is unusual in that it
changes under the action of the renormalization group.
This happens because rotations of the tubule mix two sets of fields
{--} the in-plane and out-of-plane phonons {--} having different
scaling dimensions. 
In fact, it is straightforward to show that Eq.~\ref{exact_rot} is
realized at time `l' by
\be\label{invariance}
\begin{array}{l l} u \rightarrow  &  u \cos \theta + 
e^{-(\nu-z)l} h \sin \theta + e^{-2(\nu-z)l} (\cos \theta -1)y \\
h \rightarrow   & h \cos \theta - e^{(\nu-z)l} u \sin \theta  -
e^{-(\nu-z)l} \sin \theta y \ .
\end{array} 
\ee
The above transformation is an exact symmetry of the free energy 
Eq.~\ref{free_WILS}. 
This transformation depends explicitly on $l$ and prevents
a simple construction of invariant free energies. At large $l$, however, 
we may derive an $l$-independent version. 
Define $\theta=A e^{(\nu-z)l}$ and assume that the condition 
\be\label{cond}
\nu(l)-z(l) < 0 
\ee
is satisfied. Near the fixed point, scaling relations to be derived
later show that
\be\label{cond2}
\nu-z= \frac{1}{3}(\nu-D +1)
\ee
and therefore  $\nu-z < 0$ for all  $\nu < D-1$. The physical case
$D=2$ requires $\nu<1$, which is always valid.

Eq.~\ref{invariance} is then, for large $l$,
\be\label{inv_infty}
\begin{array}{l l} u \rightarrow  & u + A h
 -\frac{1}{2}A^2 y +{\cal O}(e^{2(\nu-z)l}) \\
h \rightarrow   & h - A  y+{\cal O}(e^{2(\nu-z)l}) \ . \\
\end{array} 
\ee
The generalization of this symmetry to an arbitrary rotation involving
$\vec h$ is 
\be\label{inv_gen_infty}
\begin{array}{l l} u \rightarrow  & u + {\vec A} {\vec h}
-\frac{1}{2}{\vec A}^2 y +{\cal O}(e^{2(\nu-z)l}) \\
{\vec h} \rightarrow   & {\vec h}-{\vec A} y+{\cal O}(e^{2(\nu-z)l}) 
\ , \\
\end{array} 
\ee
which is the tubular phase version of a symmetry noted earlier in 
\cite{DGLP} for the free energy describing the large distance properties 
of the flat phase.

\subsection{The large distance free energy of the Phantom tubule}
\label{SUBSECT__phtub}

Let us apply the previous considerations to the construction of the free
energy for the large distance properties of phantom tubules 
(Eq.~\ref{freeenergy} with $b=0$).

In \cite{RT1,RT2}, the free energy
\bea\label{free_RT}
F(u,\vec h)&=&\frac{1}{2}\int d^{D-1}{\bf x}_{\perp} dy \left[
\kappa (\pary^2 \vec h)^2+t(\parp_{\alpha} \vec h)^2 \right.
\nonumber\\
&&\left.+
g_{\perp}(\parp_{\alpha} u)^2+g_y(\pary u+\frac{1}{2}(\pary \vec h)^2)^2
\right] \ \ ,
\eea
is given as that describing the right large distance properties of the TP.

The first thing to notice is that this free energy is not invariant
under the symmetry Eq.~\ref{inv_gen_infty}. The free energy with the
correct invariances is given by 
\bea\label{free_EG}
F(u,\vec h)&=&\frac{1}{2}\int d^{D-1}{\bf x}_{\perp} dy \left[
\kappa (\pary^2 \vec h)^2+t(\parp_{\alpha} \vec h)^2 \right.
\nonumber\\
&&+
g_{\perp}(\parp_{\alpha} u+\partial_{\alpha} \vec h \pary \vec h )^2
\nonumber\\
&&+\left.
g_y(\pary u+\frac{1}{2}(\pary \vec h)^2)^2
\right] \ ,
\eea
since the operator $\parp_{\alpha} u + \parp_{\alpha} {\vec h} \pary
{\vec h}$ is rotationally invariant.

It is important at this point to recall that the symmetry
Eq.~\ref{inv_gen_infty} is exact up to `irrelevant' terms, and the
coupling $g_{\perp}$ is irrelevant for all the entire range of $D$
(including $D=2$) in which the TP exists. If we therefore insist on including
irrelevant operators around the Gaussian fixed point, our free energy would
certainly contain a non-invariant term under Eq.~\ref{inv_gen_infty}, 
\bea\label{free_ALL}
F(u,\vec h)&=&\frac{1}{2}\int d^{D-1}{\bf x}_{\perp} dy \left[
\kappa (\pary^2 \vec h)^2+t(\parp_{\alpha} \vec h)^2 \right.
\nonumber\\
&&+
g_{\perp}^{(1)}(\parp_{\alpha} u+\partial_{\alpha} \vec h \pary \vec h )^2
+g_{\perp}^{(2)}(\parp_{\alpha}u)^2
\nonumber\\
&&+\left.
g_y(\pary u+\frac{1}{2}(\pary \vec h)^2)^2
\right] \ .
\eea
Indeed, this is the combination that appears, up to higher irrelevant
terms, in the general expression for the free energy Eq.~\ref{freeenergy},
as $g^{(1)}_{\perp}$ is the coupling to the  $F_{\alpha}F^{\alpha}$
operator, and $g^{(2)}_{\perp}$ is the coupling to $F^{\alpha}_{\alpha}$.

The usual strategy, nevertheless, is to keep just those operators that
define relevant directions of the Gaussian fixed point.  It is these 
directions that flow towards new infrared fixed points, unless a first 
order transition occurs. Adopting this approach the relevant free energy 
for the phantom tubule would be
\bea\label{free_REL}
F(u,\vec h)&=&\frac{1}{2}\int d^{D-1}{\bf x}_{\perp} dy \left[
\kappa (\pary^2 \vec h)^2+t(\parp_{\alpha} \vec h)^2 \right.
\nonumber\\
&&+\left.
g_y(\pary u+\frac{1}{2}(\pary \vec h)^2)^2
\right] \ ,
\eea
where $g_y$ defines a relevant direction for $D<5/2$ which terminates
in the tubular phase fixed point (TPFP) as shown in
Fig.~\ref{fig__BG}. Note that the symmetry Eq.~\ref{inv_gen_infty} 
is, indeed, preserved.

\subsection{The large distance free energy for the self-avoiding tubule}
\label{SUBSECT_satub}

Now let us return to the more physical model with 
the self-avoidance term,
\be\label{def_selfavoid}
\frac{b}{2}\int dy d^{D-1}{\bf x}_{\perp}\int dy^\prime d^{D-1}
{\bf x}_{\perp}^\prime
\delta^d(\vec r({\bf x}_{\perp},y)-\vec r({\bf x}_{\perp}^\prime,y^\prime)
 ) .
\ee
restored.

Following the discussion in subsection~\ref{SUBSECT__rotations}, 
we simplify the self-avoiding term Eq.~\ref{def_selfavoid} by
demanding invariance under the symmetry Eq.~\ref{inv_gen_infty}, 
\be\label{self_sim}
\frac{b}{2}\int dy d^{D-1}{\bf x}_{\perp}d^{D-1}{\bf x}_{\perp}^\prime 
\delta^{d-1}(\vec{h}({\bf x}_{\perp},y)-\vec{h}({\bf x}_{\perp}^\prime,y)) \ ,
\ee
with irrelevant terms dropped. 

The scaling dimension of the new perturbation Eq.~\ref{self_sim} at
the Gaussian fixed point is $\vap=3D-\frac{1}{2}-(\frac{5}{2}-D)d$. 
The $b$ coupling therefore defines a new relevant direction 
for $D$ tubules embedded in dimensions $d<d_c^{SA}$, where 
\be\label{crit_dim}
d_{c}^{SA}(D)=\frac{6D-1}{5-2D} \ .
\ee
Below the upper critical dimension $d_c^{SA}$ the Gaussian fixed point is
infrared unstable under this perturbation, and the large distance properties 
of the self-avoiding tubule are described by a new fixed point
(SAFP). This new fixed point merges with the 
Gaussian fixed point at the upper critical dimension where self-avoidance 
becomes a marginal perturbation. We therefore expect the critical 
properties of the 
self-avoiding tubule to be perturbative in $\vap$, as pointed out first 
in \cite{BG} (see Fig.~\ref{fig__BG}). 

In \cite{RT2}, however, it is claimed that this simple scenario is valid
only for tubules embedded in dimensions $d$ close to $d_c^{SA}(D)$. For any
dimension $d$ lower than $d_*$ (where $d_*<d_{c}^{SA}$), they 
argue for the existence of  a distinct fixed point, the bending rigidity
fixed point (BRFP) (see Fig.~\ref{fig__RT}). This fixed point is 
postulated to describe the actual critical
properties of the self-avoiding tubule for the regime $d < d_*$,
including the physical case of the $D=2$ tubule embedded in $d=3$.
If this scenario is true, the critical properties of the self-avoiding
tubule are not perturbative in $\vap$. Analytical predictions become then
extremely difficult, as there is no evident small perturbative parameter.

At this stage, therefore, we need to understand better the topology of the RG
flows in the case where self-avoidance is included. Let us review the 
arguments of \cite{RT2}. They consider the free energy 
Eq.~\ref{free_RT}, together with the self-avoiding term Eq.~\ref{self_sim}.
They include all relevant directions from the
Gaussian fixed point, and an irrelevant one defined by $g_{\perp}$.
They apply the infinitesimal renormalization group a la Wilson to derive 
an equation for the evolution of couplings. The crucial equation in 
their analysis is the RG flow equation for $g_{\perp}$
\be\label{RT_gperp}
\frac{d g_{\perp}}{d l}=\left[4\nu-z+D-3\right]g_{\perp} \ .
\ee
Now, as the RG is iterated starting near the Gaussian fixed point, 
$g_{\perp}$ decreases to zero
while the rescalings $\nu(l)$ and $z(l)$ flow towards their
SAFP values. 
For sufficiently small embedding dimension $d$ and large enough $l$
the sign of the $\beta$-function for $g_{\perp}$ changes sign. 
The coupling $g_{\perp}$ then flows to the BRFP $g_{\perp}^*
= \infty$ (see Fig.~\ref{fig__RT}). 
This argument can be made more quantitative.  
Under very reasonable assumptions, Eq.~\ref{RT_gperp} leads to a lower 
bound for $d_*$,
the highest embedding dimension in which the BRFP prevails,
\be\label{RT_dcrit}
d_*(D) > \frac{4D-1}{4-D} \ .
\ee
In particular, $d_*(2)>7/2>3$, so the physical tubule $(D=2,d=3)$
is, according to \cite{RT2}, described by the BRFP.

It is apparent that the operator
\be\label{oper_alphau}
\partial_{\alpha} u \partial^{\alpha} u \ 
\ee
plays a fundamental role in this argument. Let us examine it more closely.
In an expansion in irrelevant operators around the Gaussian fixed point, 
it appears
in two ways, which we labeled $g_{\perp}^{(1)}$ and $g_{\perp}^{(2)}$ in
Eq.~\ref{free_ALL}. 

First of all, the operator associated to $g_{\perp}^{(1)}$ is
invariant under the symmetry Eq.~\ref{inv_gen_infty}, as it appears in
the invariant combination 
\be\label{partialu_inv}
\partial_{\alpha}u+\partial_{\alpha}\vec h\partial_{y} \vec h \ .
\ee
In contrast $g_{\perp}^{(2)}$ couples to a subdominant piece of the 
operator $F^{\alpha}_{\alpha}\equiv \partial_{\alpha} u \partial^{\alpha}u+ 
\partial_{\alpha} \vec h \partial^{\alpha} \vec h $ (see
Eq.~\ref{inv_op}). In fact, from our earlier symmetry arguments, it is
suppressed by a factor ${\cal O}(e^{2(\nu-z)})$ with respect to
the dominant piece $(\partial_\alpha \vec h)^2$ which couples to the
marginal direction $t$.
Provided $\nu-z<0$ the coupling $g_{\perp}^{(2)}$ is thus irrelevant
and may be dropped from the free energy.

We have argued that the most general free energy dictating the large
distance properties of the tubule is given by Eq.~\ref{free_EG}
together with self-avoidance (Eq.~\ref{self_sim}). 
For $g_{\perp}$ vanishing, the infrared stable fixed point of the
theory is the SAFP. The key issue is now whether this fixed point is
stable with respect to perturbations by $g_{\perp}$. 
Since the properties of the SAFP are perturbative 
in $\vap$, the same applies to the critical exponents.
Experience with typical multicritical behavior suggest that we should
not expect the exponent associated with the $g_{\perp}$ direction 
to change so much from its gaussian value $3/2-D$ that it changes
sign \cite{AM}.

In conclusion, the simplest free energy describing the large distance 
properties of the self-avoiding tubule is given by
\bea\label{free_self_EG}
F(u,\vec h)&=&\frac{1}{2}\int d^{D-1}{\bf x}_{\perp} dy \left[
\kappa (\pary^2 \vec h)^2+t(\parp_{\alpha} \vec h)^2+
g_y(\pary u+\frac{1}{2}(\pary \vec h)^2)^2
\right] \nonumber\\
&&+
\frac{b}{2}\int dy d^{D-1}{\bf x}_{\perp}d^{D-1}{\bf x}_{\perp}^\prime 
\delta^{d-1}(\vec{h}({\bf x}_{\perp},y)-\vec{h}({\bf x}_{\perp}^\prime,y)) \ .
\eea
This is the starting point of all our subsequent analysis.

\section{The Scaling Relations}\label{SECT__scalrel}

Having identified the right free energy, we turn now to the derivation 
of the different critical exponents of the theory. We use
the conventional field theory formalism \cite{AM}, following \cite{BG}.

The scaling dimensions of the fields and coordinates are
$[y]=1$,$[x_{\perp}]=2$, $[\vec h]=\frac{5}{2}-D$ and
$[u]=4-2D$. This implies
\be\label{engin_dim}
[b]=-\vap \ , \ [g_y]=2D-5 \ ,
\ee
with
\be\label{eps_dim}
\vap=3D-\frac{1}{2}-\frac{5-2D}{2}d .
\ee
Following the arguments in \cite{BG}, one can show that
the free energy Eq.~\ref{free_self_EG} renormalizes onto itself with
\bea\label{free_REN_SA}
F(u,\vec h)&=&\frac{1}{2}\int d^{D-1}{\bf x}_{\perp}^R dy \left[
Z \kappa (\pary^2 \vec h^R)^2+Z_{\perp}t(\parp_{\alpha}^R \vec h^R)^2
\right.
\nonumber\\
&&\left.+
g_y^R \mu^{-5+2D}(\pary u^R+\frac{1}{2}(\pary \vec h^R)^2)^2
\right]  
\\\nonumber
&&+\frac{b^R Z_b \mu^{\vap}}{2}\int d^{D-1}{\bf x}_{\perp}^R d^{D-1}
{\bf x}_{\perp}^{R}dy \delta^{d-1}(\vec h^R({\bf x}_{\perp}^R,y)-
\vec h^R({\bf x}_{\perp}^{\prime R},y)) , 
\eea
where the Ward identity implied by Eq.~\ref{inv_gen_infty} is used so
that there is no independent wave function renormalization for the
field $u$. Furthermore, it is not difficult to show, using the MOPE formalism
\cite{MOPE2}, that the bending rigidity is not renormalized so $Z=1$, as
first pointed out in \cite{BG}. Thus we have
\bea\label{bare_ren}
\vec h^R({\bf x}_{\perp}^R,y)&=&Z_{\perp}^{\frac{1-D}{4}} 
\vec h ({\bf x}_{\perp},y)
\nonumber\\
{\bf x}_{\perp}^R&=&Z^{\frac{1}{2}}_{\perp} {\bf x}_{\perp}
\\\nonumber
b^R&=&b\mu^{-\vap}Z^{-1}_b Z^{(1-D)\frac{d+3}{4}}
\\\nonumber
g_y^R&=&\mu^{5-2D}g_yZ_{\perp}^{\frac{D-1}{2}} \ .
\eea

Using these definitions we will consider two correlators, which enable
us to determine the exponents of the theory. In the original paper
\cite{BG}, the correlator
\be\label{BG_corrh}
G_h({\bf x}_{\perp},y)\equiv -\frac{1}{2(d-1)}\left\langle 
(\vec h({\bf x}_{\perp},y) -\vec h({\bf 0},0) )^2 \right\rangle 
\ee
was considered as well the correlator involving the $u$ fields,
\be\label{BG_corru}
G_u({\bf x}_{\perp},y)\equiv \left\langle \pary u({\bf x}_{\perp},y) 
\pary u({\bf 0},0)
\right\rangle \ .
\ee

At the fixed point, the first correlator satisfies
\be\label{FP_eq}
\left\{\mu\frac{\partial}{\partial \mu}+\frac{\delta}{2} {\bf x}_{\perp}
\frac{\partial}{\partial {\bf x}_{\perp}}+\frac{D-1}{2}\delta \right\}
G^R_h({\bf x}_{\perp},y)=0 \ ,
\ee
which, combined with simple scaling law
\be\label{FP_sca_h}
\left\{\mu\frac{\partial}{\partial \mu}-y\frac{\partial}{\partial y}
-2{\bf x}_{\perp} \frac{\partial}{\partial {\bf x}_{\perp}}+(5-2D) \right\}
G^R_h({\bf x}_{\perp},y)=0 \ ,
\ee
gives us the fixed point renormalization group equation,
\be\label{RG_eq1}
\left\{y\frac{\partial}{\partial y}+\frac{1}{z} {\bf x}_{\perp}
\frac{\partial}{\partial {\bf x}_{\perp}}-2\zeta \right\}
G^R_h({\bf x}_{\perp},y)=0 \ .
\ee

A renormalization group equation may also be derived for $G_u$. To do so
we must use once again the Ward identity that fixes the wave
function renormalization for $u$. Eq.~\ref{FP_eq} is now,
\be\label{FP_eq_u}
\left\{\mu\frac{\partial}{\partial \mu}+\frac{\delta}{2}{\bf x}_{\perp}
\frac{\partial}{\partial {\bf x}_{\perp}}+(D-1)\delta \right\}
G^R_u({\bf x}_{\perp},y)=0 \ ,
\ee
Eq~\ref{FP_sca_h} reads for the $u$ case,
\be\label{FP_sca_u}
\left\{\mu\frac{\partial}{\partial \mu}-y\frac{\partial}{\partial y}
-2{\bf x}_{\perp} \frac{\partial}{\partial{\bf x}_{\perp}}+(6-4D) \right\}
G^R_u({\bf x}_{\perp},y)=0 \ ,
\ee
leading finally to
\be\label{RG_eq2}
\left\{y\frac{\partial}{\partial y}+\frac{1}{z} {\bf x}_{\perp}
\frac{\partial}{\partial {\bf x}_{\perp}}-2\zeta_u \right\}
G^R_h({\bf x}_{\perp},y)=0 \ ,
\ee
where 
\bea\label{all_scaling}
\delta&=& \left. \mu\frac{d}{d \mu} \right|_0 log Z_{\perp} 
\nonumber\\
z&=&\frac{2}{4+\delta} 
\\
\zeta&=&\frac{5-2D}{2}+\frac{1-D}{4}\delta
\nonumber\\\nonumber
\zeta_u&=&1+\frac{1-D}{z} \ .
\eea
Both, Eqs~\ref{RG_eq1} and \ref{RG_eq2} may be solved explicitly, yielding 
\bea\label{RG_sol}
G_h({\bf x}_{\perp},y)&=&y^{2\zeta}F_1(\frac{y}{|{\bf x}_{\perp}|^z})
=|{\bf x}_{\perp}|^{\nu}F_2(\frac{y}{|{\bf x}_{\perp}|^z})
\\\nonumber
G_u(x_{\perp},y)&=&y^{2\zeta_u}F_1^\prime(\frac{y}{|{\bf x}_{\perp}|^z})
=|{\bf x}_{\perp}|^{2\frac{\zeta_u}{z}} 
F_2^\prime(\frac{y}{|{\bf x}_{\perp}|^z}) 
\ ,
\eea
where $\nu=\frac{\zeta}{z}$.
Transforming Eq.~\ref{RG_sol} to momentum space gives
\bea\label{RG_sol_mom}
G_h({\bf p}_{\perp},q)^{-1}&=&|{\bf p}_{\perp}|^{2+\eta_{\perp}}
f(\frac{q}{|{\bf p}_{\perp}|})
\\\nonumber
G_u({\bf p}_{\perp},q)^{-1}&=&|{\bf p}_{\perp}|^{z\eta_u}h(\frac{q}
{|{\bf p}_{\perp}|})
\eea
with
\bea\label{scal_mom_law}
\eta_{\perp}&=&-2+4z
\nonumber\\
\eta_u&=&\frac{2 \nu}{z} \ .
\eea
These scaling laws were first derived in \cite{BG} and \cite{RT1} 
respectively.

We conclude that all the critical exponents of our Free energy 
Eq.~\ref{free_self_EG} at any putative fixed point
may be expressed in terms of a single parameter, say $\delta$. The
task of computing critical exponents translates into the task of
evaluating $\delta$ at the corresponding fixed point.

\section{The Phantom tubule}\label{SECT__phtub}

The theoretical considerations in subsection~\ref{SUBSECT__phtub}
lead us to consider Eq.~\ref{free_REL} as the right free energy 
describing the large distance properties of the phantom tubule.
In fact they allow us to solve the phantom tubule 
phase exactly, simply by performing the shift
\be\label{shift}
u \rightarrow u^\prime-\frac{1}{2} \int_{0}^y dz(\partial_z \vec{h})^2 \ ,
\ee
where the lower bound for the integral is arbitrary and corresponds to
translations of the zero mode. The free energy Eq.~\ref{free_REL}
is then a sum of Gaussian terms. Let us compute the anomalous elasticity,
determined by the correlator Eq.~\ref{BG_corru}
\be\label{def_corr}
G_u({\bf x}-{\bf z})=\langle \pary u({\bf x}) \pary u({\bf z}) \rangle 
\ , \ G_u({\bf p})=\int d({\bf x-z}) e^{-{\bf p}({\bf x}-{\bf z})}
G({\bf x}-{\bf z}) .
\ee
The elasticity constant is given by
\be\label{g_elastic}
g_{y}({\bf p})=\frac{1}{G_u({\bf p})}  \ .
\ee
At tree level $g_{y}({\bf p})=g_{y}$.

The general case amounts to performing the shift Eq.~\ref{shift}.
Equivalently the loop expansion may be performed to all orders. 
The diagrams consists of a necklace of $\vec h$ loops which can be 
resummed, yielding 
\bea\label{g_exact}
\frac{1}{g_{y}({\bf p})}&=&\frac{1}{g_y}+
\frac{d-1}{2}\int d^{D-1}\hat{{\bf q}}_{\perp} d \hat{q_y}
\frac{q^2_{y}(p-q)^2_{y}}{(\kappa q^4_y+tq^2_{\perp})
(\kappa (q_y-p_y)^4+t({\bf q}_{\perp}-{\bf p}_{\perp})^2)}
\nonumber\\
&+&
\left(\frac{d-1}{2}\int d^{D-1}\hat{{\bf q}}_{\perp} d \hat{q_y}
\frac{q^2_{y}}{\kappa(q^2_y)^2+t{\bf q}^2_{\perp}}\right)^2 
\delta({\bf p}) \ ,
\eea
which for ${\bf p} \ne 0$ is
\be\label{g_explicit}
g_{y}({\bf p})=\frac{g_y}{1+\frac{(d-1) g_y}{2 t^2} f(D-1) p_y^{2D-5}
C(\frac{p_y}{|{\bf p}_{\perp}|^z})} \ ,
\ee
where $f(d)=\int d^d\hat{t}\frac{1}{(t^2+1)^4}$, 
\be\label{scaling}
C(y)=\int^{+\infty}_{-\infty}d \hat{z} \int^1_0 d x 
\left(\frac{x(1-x)}{y^4}+\frac{\kappa}{t}(x(1-z)^4+(1-x)z^4)\right)^
{\frac{D-5}{2}},
\ee
and the exponent $z$ is
\be\label{z-exponent}
z=1/2 \ .
\ee

Recall that Eq.~\ref{g_explicit} is valid for any value of $g_y$.
The Gaussian fixed point ($g_y=0$) is unstable to perturbations along this
direction, and the coupling $g_y$ is driven to $g_y=\infty$ in the
infrared, which is the fixed point describing the physics of the phantom tubule
(PTFP). At the PTFP $g_y({\bf p})$ has the form
$p_y^{\eta_u}g(\frac{p_y}{|{\bf p}_{\perp}|^z})$, as predicted by 
Eq.~\ref{RG_sol} at any fixed point.

We can easily recover now the results in \cite{RT1,RT2} from our exact
solution Eq.~\ref{g_explicit} and Eq.~\ref{scaling} at the PTFP:
\begin{itemize}
\item{ For $y\equiv \frac{p_y}{|{\bf p}_{\perp}|^z} \rightarrow \infty$ }

we have
\be\label{c_infty}
C(y)_{y \rightarrow \infty} \sim (\frac{\kappa}{t})^{\frac{D-5}{2}}
\frac{2}{D-3}\int^{+\infty}_{-\infty} d \hat{z} z^2(1-z)^2
\frac{(1-z)^{2(D-3)}-z^{2(D-3)}}{(1-z)^4-z^4}
\ee
which converges, both in the infrared and the ultraviolet for
$\frac{3}{2} < D < \frac{5}{2} $ (the dimensions in which the tubular
phase exists).
For small $p_y$ 
\be\label{g_infty}
g_{y}({\bf p}) \sim p_y^{\eta_{u}} \ , \ \eta_{u}=5-2D .
\ee

\item{ For $y\equiv \frac{p_y}{|{\bf p}_{\perp}|^z} \rightarrow 0$ }

we have

\be\label{c_zero}
C(y)_{y\rightarrow 0}\sim \frac{y^{-2D+5}}{2}(\frac{t}{\kappa})^D
\int^1_0(x(1-x))^{\frac{2D-5}{2}}
\int^{+\infty}_{0} d \hat{z} z^{1/4}(1+z)^{\frac{D-5}{2}}
\ee
so $C(y)_{\rightarrow 0} \sim y^{\eta_u} Const.$ where $Const.$ is
a convergent integral for $\frac{3}{2} < D < \frac{5}{2} $.
For small $p_{\perp}$ at $D=2$
\be\label{g_zero}
g_y({\bf p}) \sim p_{\perp}^{1/2} \ .
\ee
\end{itemize}

To conclude let us connect with the results in 
section~\ref{SECT__scalrel}. At the PTFP we have $\delta=0$, 
and rest of the exponents from this result and scaling relations.

\section{The self-avoiding tubule}
\label{SECT__SAtub}

We have argued that Eq.~\ref{free_self_EG} is the appropriate
free energy to consider once self-avoidance is included. 
The task of computing the critical exponents of this theory is not easy,
since from \cite{BG} we know that the results of the
$\vap(d)$-expansion are not very robust to higher order perturbations.

In order to get an estimate for the exponents at the SAFP, we 
compute the critical exponents to lowest order in $\vap(D,d)$ for 
arbitrary internal dimension $D$. Existing techniques
\cite{HWA1,DAVWI} then allow us to perform more sophisticated extrapolations
which produce reliable estimates for the critical exponents.

\subsection{The computation of $\delta$}\label{SUBSECT__delta}

We follow the MOPE formalism \cite{MOPE2}, employing dimensional
regularization and minimal subtraction, used first in this problem 
in \cite{BG}, to compute the $\delta$ exponent.

Within the MOPE formalism, one may prove that the free energy 
Eq.~\ref{free_self_EG} renormalizes onto itself.
It also identifies the diagrams to compute that yield the RG functions 
determining the critical exponents. For details on this formalism
we refer to \cite{MOPE1,MOPE2}, and for its implementation in the tubular 
case, we refer to the original BG \cite{BG} calculation.

\begin{figure}[htb]
\epsfxsize=5 in \centerline{\epsfbox{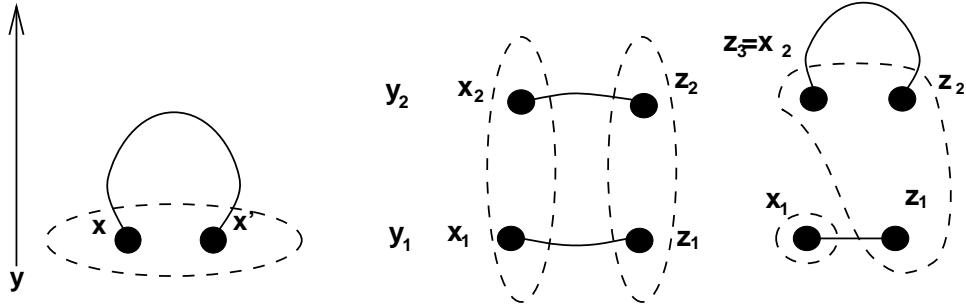}}
\caption{One loop diagrams contributing to the renormalization of the
free energy.}
\label{fig__diag}
\end{figure}

The first step is to compute the two-point function $G_h$ for arbitrary 
$D$ at $b=0$. The result is
\bea\label{two_point}
G_h^0({\bf x}_{\perp},y)&=&\mbox{}-\frac{|{\bf x}_{\perp}|^{2-D}}
{(\frac{5}{2}-D)(2\pi)^{\frac{D+1}{2}}}
\left[|{\bf x}_{\perp}|^{\frac{1}{2}}\int^{+\infty}_{0} dt t^{\frac{D}{2}-1}
K_{\frac{1-D}{2}}(t)\cos(t^{\frac{1}{2}}w)\right.
\nonumber\\
&&\left.+\frac{y}{2}\int^{+\infty}_{0} dt t^{\frac{D-3}{2}}
K_{\frac{3-D}{2}}(t)\sin(t^{\frac{1}{2}}w)\right]  
\eea
where $w=\frac{y}{|{\bf x}_{\perp}|^{\frac{1}{2}}}$ and $K_{\nu}$ is a modified
Bessel function. There are two particular cases of interest. At $y=0$ we have
\be\label{y_zero}
G_h^0({\bf x}_{\perp},0)=-\frac{|{\bf x}_{\perp}|^{\frac{5}{2}-D}
\Gamma(\frac{1}{4})\Gamma(\frac{D}{2}-\frac{1}{4})}
{(\frac{5}{2}-D) \pi^{\frac{D+1}{2}} 2^{\frac{5}{2}}} \ .
\ee
For the physical value $D=2$ it follows from 
$K_{\frac{1}{2}}(t)=K_{-\frac{1}{2}}(t)=
\left(\frac{\pi}{2t}\right)^{1/2}e^{-t}$, that 
\be\label{y_zero_2d}
G_h^0(x,y)=-\frac{|x|^{\frac{1}{2}}}{2 \pi^{\frac{1}{2}}} 
e^{-\frac{w^2}{4}}-\frac{y}{4} {\rm erf}(\frac{w}{2}) \ ,
\ee
where ${\rm erf(x)}$ denotes the error function.
This result is in complete agreement with that quoted in \cite{BG}.
The next step is to perform the MOPE for the operator
\be\label{delta_fun}
\phi\{{\bf x}_{\perp},{\bf x}_{\perp}^\prime,y\}=\delta^{(d-1)}(\vec
h({\bf x}_{\perp},y)-\vec h({\bf x}_{\perp}^\prime,y)) \ .
\ee
This is easily done using standard techniques, with the result
\be\label{mope_delta}	
\phi\{{\bf x}_{\perp},{\bf x}_{\perp}^\prime,y\}=C^1_{\phi}({\bf
x}_{\perp}-{\bf x}_{\perp}^\prime)+C^{\alpha \beta}_{\phi}({\bf
x}_{\perp}-{\bf x}_{\perp}^\prime)
:\nabla_{\alpha} \vec h({\bf x}_{\perp}^0,y) \nabla_{\beta} \vec
h({\bf x}_{\perp}^0,y): + \cdots \ ,
\ee
where ${\bf x}_{\perp}^0=\frac{{\bf x}_{\perp} + {\bf x}_{\perp}^{\prime}}{2}$, and the Wilson coefficients are
\be\label{first_coef}
C_{\phi}({\bf u})=\frac{1}{(4\pi)^{\frac{d-1}{2}}
(-G_h^0({\bf u},0))^{\frac{d-1}{2}}} \ , \
C_{\phi}^{\alpha \beta}({\bf u})=-\frac{u^{\alpha} u^{\beta}}
{4(4\pi)^{\frac{d-1}{2}}(-G_h^0({\bf u},0))^{\frac{d+1}{2}}} \ .
\ee
We also need the MOPE for the product of two of these operators.
One finds 
\be\label{mope_two}
\phi\{{\bf x}^{\perp}_1,{\bf z}^{\perp}_1,y_1\}\phi\{{\bf
x}^{\perp}_2,{\bf z}^{\perp}_2,y_2\}=C^{\phi}_{\phi \phi}({\bf
x}^{\perp}_1-{\bf x}^{\perp}_2,
{\bf z}^{\perp}_1-{\bf z}^{\perp}_2,y_1-y_2)\phi\{{\bf x}_{\perp},{\bf
z}_{\perp},y\} + \cdots \ ,
\ee
where ${\bf x}_{\perp}=\frac{{\bf x}^{\perp}_1+{\bf x}^{\perp}_2}{2}$, 
$y=\frac{y_1+y_2}{2}$ and
${\bf z}_{\perp}=\frac{{\bf z}^{\perp}_1+{\bf z}^{\perp}_2}{2}$ and 
\be\label{wils_coef}
C^{\phi}_{ \phi \phi}({\bf u},{\bf v},w)=\frac{1}{(4\pi)^{\frac{d-1}{2}}}
\frac{1}{(-G_h^0({\bf u},w)-G_h^0({\bf v},w))^{\frac{d-1}{2}}} \ .
\ee
This is all we need to compute the critical exponents. This MOPE
corresponds to the diagrams in Fig.~\ref{fig__diag}. The last diagram 
for the renormalization of $b$ is not necessary to compute, 
as it cancels against the renormalization of $Z_{\perp}$.

Expanding the renormalized action Eq.~\ref{free_REN_SA}, and using the 
MOPE Eq.~\ref{mope_delta},
\bea\label{first-ren}
&& -\frac{b^R\mu^{\vap}}{2}\int d^{D-1} {\bf x}_{\perp} d^{D-1}
{\bf x}_{\perp}^\prime dy
\phi\{{\bf x}_{\perp},{\bf x}_{\perp}^\prime,y\}=
\\\nonumber &&
-\frac{b^R\mu^{\vap}}{2} \left( \int 
d^{D-1} {\bf x}_{\perp} d^{D-1} {\bf x}_{\perp}^\prime dy
C^1_{\phi}({\bf x}_{\perp}-{\bf x}_{\perp}^\prime)\right.+
\\\nonumber && \left.+\int 
d^{D-1} {\bf x}_{\perp} d^{D-1} {\bf x}_{\perp}^\prime dy
C^{\alpha,\beta}_{\phi}({\bf x}_{\perp}-{\bf x}_{\perp}^\prime) 
\nabla_{\alpha}\vec h({\bf x}_{\perp}^0,y) \nabla_{\beta} \vec h
({\bf x}_{\perp}^0,y) +\cdots
\right) \ .
\eea
The first term of Eq.~\ref{first-ren} renormalizes
the identity operator and therefore may be neglected in computing expectation
values of operators. The second term determines $Z_{\perp}$, with
the result,
\be\label{Z_perp}
Z_{\perp}=1+\frac{1}{D-1}(\frac{5}{2}-D)^{\frac{2D+2}{5-2D}}
\frac{{\pi}^{\frac{3D+3}{10-4D}}2^{\frac{11-3D}{5-2D}}}
{\left(\Gamma(\frac{1}{4})\Gamma(\frac{D}{2}-\frac{1}{4})\right)^
{\frac{2D+2}{5-2D}}}\frac{b^R}{\Gamma(\frac{D-1}{2})}\ \frac{1}{\vap} \ .
\ee
At one loop there is no renormalization for $Z$, a result that 
is also true at any order in perturbation theory \cite{BG}.

Expanding the $\delta$-function, and performing the MOPE in 
Eq.~\ref{mope_two}, we find
\bea\label{zb_ren1}
&&\frac{(b^{R}\mu^{\vap})^2}{8}\int d^{D-1}{\bf x}_{\perp}^0
d^{D-1}{\bf x}_{\perp}^{0\,\prime} dy_0
\delta^{d-1}(\vec h({\bf x}_{\perp}^0,y_0)-
\vec h({\bf x}_{\perp}^{0\,\prime},y_0) 
\nonumber\\&&
\times \int d^{D-1}{\bf z} 
d^{D-1}{\bf w} dy C^{\phi}_{\phi \phi}({\bf z},{\bf w},y) \ ,
\eea
where higher terms in the MOPE Eq.~\ref{mope_two} are neglected
as they do not give rise to poles in $\vap$. To find 
$Z_{b}$, we must compute the last integral in Eq.~\ref{zb_ren1}.
This is done by performing the angular integration and then 
changing variables to $u=\frac{|{\bf z}|^{1/2}}{y}$ and 
$v=\frac{|{\bf w}|^{1/2}}{y}$.
The result is
\bea\label{zb_int}
\int d^{D-1}{\bf z} d^{D-1}{\bf w} dy C^{\phi}_{\phi \phi}({\bf
z},{\bf w},y)&=&
\frac{8}{(4 \pi)^{\frac{d-1}{2}}}
\left(\frac{2 \pi^{\frac{D-1}{2}}}{\Gamma(\frac{D-1}{2})}\right)^
{\frac{d-1}{2}} 
\\\nonumber&& \times
\left((\frac{5}{2}-D)(2 \pi)^{\frac{D+1}{2}}\right)^{\frac{d-1}{2}}
\int_0^{\frac{1}{\mu}}dy y^{\vap-1}
\\\nonumber&& \times \int_0^{\frac{1}{\mu y}}du
\int_0^{\frac{1}{\mu y}} dv \frac{u^{2D-3}v^{2D-3}}{(f(u)+f(v))^
{\frac{d-1}{2}}} \ ,
\eea
where 
\bea\label{define_fu}
f(u)&=&u^{4-2D}\left\{u\int^{+\infty}_{0} dt t^{\frac{D}{2}-1}
K_{\frac{1-D}{2}}(t)\cos(\frac{t^{1/2}}{u})\right.
\nonumber\\&&\left.
+\frac{1}{2}\int^{+\infty}_0dt t^{\frac{D-2}{2}}
K_{\frac{3-D}{2}}(t)\sin(\frac{t^{1/2}}{u})\right\} \ .
\eea
Using 
\bea\label{f_identity}
&&
\int_0^{\frac{1}{\mu y}}du
\int_0^{\frac{1}{\mu y}} dv \frac{u^{2D-3}v^{2D-3}}{(f(u)+f(v))^
{\frac{d-1}{2}}} 
\\\nonumber
&=&
\int_0^{+\infty} du
\int_0^{+\infty} dv \frac{u^{2D-3}v^{2D-3}}{(f(u)+f(v))^
{\frac{d-1}{2}}} + \rho(y) \ ,
\eea
where $\rho(y)$ is a continuous function that vanishes at $y=0$,  
and adding a factor of $2$  corresponding to the two ways one
can perform the MOPE in the diagram in Fig.~\ref{fig__diag}, we get finally
\be\label{zb_renorma}
Z_{b}=1+\frac{(\frac{5}{2}-D)^{\frac{4D-3}{5-2D}}
2^{\frac{(4D-3)(D-3)}{2(5-2D)}+4}}
{\pi^{\frac{-7(D-1)}{2(5-2D)}} \Gamma(\frac{D-1}{2})^2} I(D) 
\frac{b^R}{\vap} \ ,
\ee
with 
\be\label{def_ID}
I(D)= \int_0^{+\infty} du
\int_0^{+\infty} dv \frac{u^{2D-3}v^{2D-3}}{(f(u)+f(v))^
{\frac{4D-3}{5-2D}}} \ .
\ee
We have thus succeeded in renormalizing the theory at the one loop
level. The evaluation of this integral is discussed in the Appendix. 
The next step is to compute the exponent $\delta$.
We begin with the computation of the $\beta$ function. There
are two of them. Defining $a_1$ and $b_1$ via 
$Z_{\perp}=1+\frac{b^R}{\vap}a_1$
and $Z_{b}=1+\frac{b^R}{\vap}b_1$, we have
\be\label{beta_fun}
\beta_b(b^R)=-\vap b^R+(a_1+\frac{7-\vap}{2(5-2D)}b_1)(b^{R})^2 \ ,
\ee
and
\be\label{beta_g}
\beta_{g_y}(b^R,g_y^R)=(5-2D+\frac{D-1}{2}\Delta(b^R))g^R_y \ , 
\ee
where $\Delta(b^R)=\mu\frac{d}{d \mu} log Z_{\perp}$.
There is a nontrivial fixed point (the SAFP) at 
\be\label{BGFP_D}
b^{R*}=\frac{\vap}{a_1+\frac{7-\vap}{2(5-2D)}b_1} \ ,
\ee  
and, for $D<\frac{5-\Delta(b^{R*})/2}{2-\Delta(b^{R*})/2}$
\be\label{BGFP_Dg}
g_y^R=+\infty \ .
\ee
Using 
\be\label{delta_def}
\delta=\Delta(b^{R*})=\left. \mu\frac{d}{d \mu} log Z_{\perp}
\right|_{b^R=b^{R*}}
\ee
we obtain the final result 
\be\label{gen_res}
\delta=-\frac{\vap}{\frac{7}{2(5-2D)}+\frac{a_1}{b_1}} \ .
\ee
Plugging Eq.~\ref{Z_perp} and Eq.~\ref{zb_renorma} into Eq.~\ref{gen_res}
yields  
\be\label{delta_loop}
\delta=-\frac{(\frac{5}{2}-D)}
{\frac{7}{4}+ \vartheta(D) I(D)}\vap \ ,
\ee
where
\be\label{mess}
\vartheta(D)=(D-1)\frac{(\Gamma(\frac{1}{4})\Gamma(\frac{D}{2}-\frac{1}{4}))^
{\frac{2D+2}{5-2D}}2^{\frac{2D^2-25D/2+27/2}{5-2D}}}
{\pi\Gamma(\frac{D-1}{2})} \ .
\ee
The scaling relations after Eq.~\ref{RG_eq2} and 
Eq.~\ref{scal_mom_law} determine the rest of the critical exponents
provided we have a good determination of $\delta$. 
As only the first term in the $\vap$ expansion is available
this will require refined methods to improve the perturbative
expansion. This is an involved subject to which we now turn.

\subsection{Analysis of the results}\label{SUBSEC__results}

\begin{figure}[htb]
\epsfxsize=5in \centerline{\epsfbox{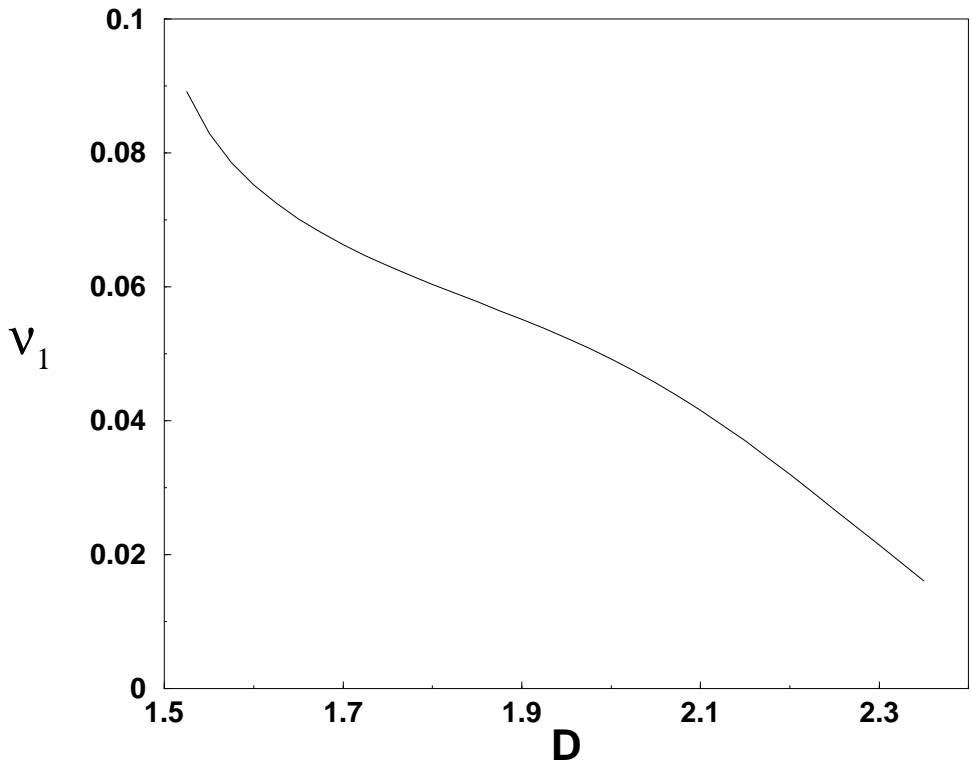}}
\caption{Plot of $\nu_1(D)$ as a function of $D$.}
\label{fig__nu1}
\end{figure}

From Eq.~\ref{delta_loop} and the scaling relations 
we get explicit forms for the critical exponents of the self-avoiding
tubular phase. For example, the radius of gyration or size exponent 
$\nu$ reads,
\be\label{nu_loop}
\nu(D)=\frac{5-2D}{4}+\nu_1(D)\vap(D,d)+\cdots \ ,
\ee
where $\nu_1(D)$ is plotted as a function of $D$ in Fig.~\ref{fig__nu1}.

\begin{figure}[hp]
\epsfxsize=5in \centerline{\epsfbox{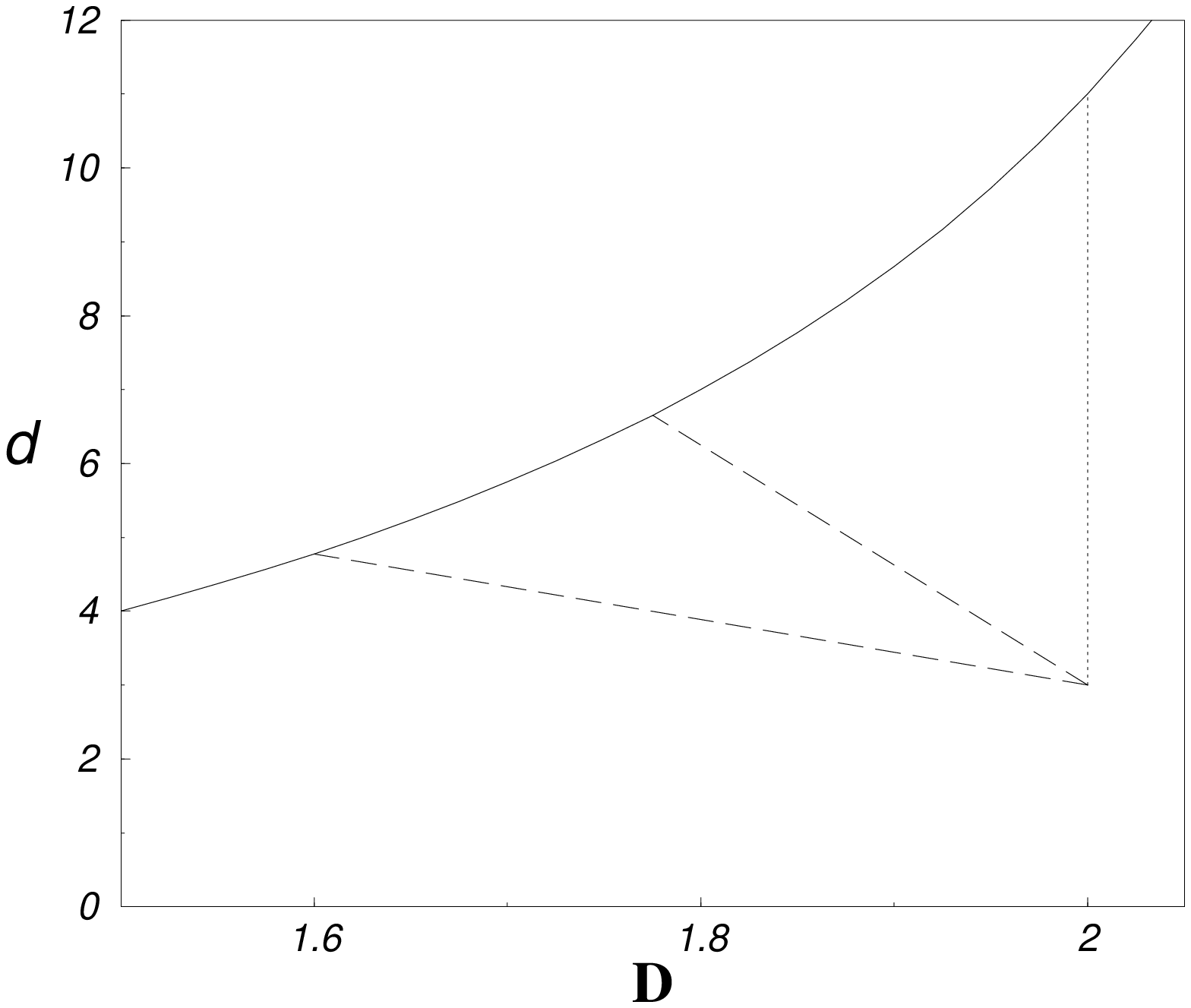}}
\caption{The solid line is the $\vap=0$ curve. The naive $\vap$ expansion 
is marked with a dotted line. Other possible expansions from the physical
interesting case $(D=2,d=3)$, are marked with dashed lines.}
\label{fig__Ddplot}
\end{figure}

As already noticed in \cite{BG}, a direct application of Eq.~\ref{nu_loop} 
to a physical membrane is not robust with respect to second order 
corrections. This is a consequence of the point $(D=2,d=3)$ being too far 
from the point $(2,11)$ on the critical curve $\vap=0$. One can try,
instead, to perform a generalized $\vap$ expansion around any other
point on the critical curve $(D_0,d_0=\frac{6D_0-1}{5-2D_0})$ (see
Fig.~\ref{fig__Ddplot}) and hope to find a new expansion
in which the corrections to Eq.~\ref{nu_loop} are minimized. In this case 
one may expect reliable one loop results.

As an example, let us rewrite Eq.~\ref{nu_loop} in terms of $D-D_0$,
and keep the leading terms,
\bea\label{nu_double_vap}
\nu(D)&=&\frac{5-2D_0}{4}-\frac{D-D_0}{2}+\nu_1(D_0+D-D_0)\vap(D,d)
+{\cal O}(\vap^2)
\nonumber\\
&=&\frac{5-2D_0}{4}-\frac{D-D_0}{2}+
\nu_1(D_0)\vap(D,d)
\\\nonumber
&& +\, {\cal O}(\vap^2,\vap(D-D_0),(D-D_0)^2) \ .
\eea
One can expand around any point $(D_0,d_0)$ in the
$\vap=0$ curve, but at the expense of dealing with the double
expansion in $\vap$ and $D$ in Eq.~\ref{nu_double_vap}.
Furthermore, critical quantities depend, in principle, on a new 
parameter $D_0$, but should obviously be 
independent of it. There are established techniques to select the best 
$D_0$, such as the minimal sensitivity method of Hwa \cite{HWA1}.
Anyway, the expansion in $\vap,D$ is just a particular case of a more
general situation \cite{DAVWI}, as we may choose any new set of variables
$\{ x(D,\vap),y(D,\vap) \}$ and re-express Eq.~\ref{nu_double_vap} as an
expansion around the critical curve 
$(x_0=x(D_0,0),y_0=y(D_0,0))$,
\be\label{nu_xy}
\nu(D,\vap)=\nu(x,y)=\nu(x_0,y_0)+\Delta x \nu_{1,0}(x_0,y_0)+
\Delta y \nu_{0,1}(x_0,y_0)+\cdots
\ee
where $\Delta x=x(D,\vap)-x_0$, $\Delta y=y(D,\vap)-y_0$.

Now the goal is to choose a good set of variables $\{x,y\}$ 
from the infinite number of possibilities. 

A good expansion variable must meet two basic requirements;
\begin{enumerate}
\item Independence of $D_0$, at least in some interval of $D_0$s.
\item Within this region, results must be quantitatively correct.
\end{enumerate}

The way we choose to meet requirement 1, is to look for $\nu$ to
plateau within some given accuracy for an interval of $D_0$. 
To meet requirement 2, we will systematically
study the $(2,d)$ line, starting close to the $d=11$ where even the
naive $\vap$ expansion is expected to work and agree well with the
Flory estimate. At large $d$ a good expansion variable should exhibit 
a broad plateau around the Flory result. At this stage, the expansions
that do not deliver accurate enough results cannot be trusted, 
and should therefore be rejected.
As $d$ is decreased we check that the plateau remains stable for the
good expansion variables, so that we have a reliable extrapolation
at $d=3$. In addition, we expect reasonable agreement with the Flory estimate.

We examine the $\nu$ exponent as well as the $\zeta$ exponent using 
the above techniques.  The reason for analyzing both
exponents is that they have different dependencies on $\delta$, as expressed in
Eq.~\ref{all_scaling}. This may result in different expansion variables
being appropriate for different exponents. Of course, results must 
be consistent with the scaling 
relations expressed by Eq.~\ref{all_scaling}.

In this paper, we implement Hwa minimal sensitivity scheme \cite{HWA1}, and
we explore the following distinct expansions;
\begin{itemize}
\item expansion A: $\{x=D, y=\vap \}$.
\item expansion B: $\{x=D,y=d \}$. 
\item expansion C: $\{ x=D, y=D_0(d)=\frac{5d+1}{2(3+d)} \}$.
\item expansion D: $\{x=\vap,y=D_0(d)\}$. 
\end{itemize}
which have been previously used in a different context in \cite{DAVWI}.

\subsubsection{Corrections to mean field}

\begin{figure}[hp]
\centerline {\epsfxsize = 2.5in \epsfbox{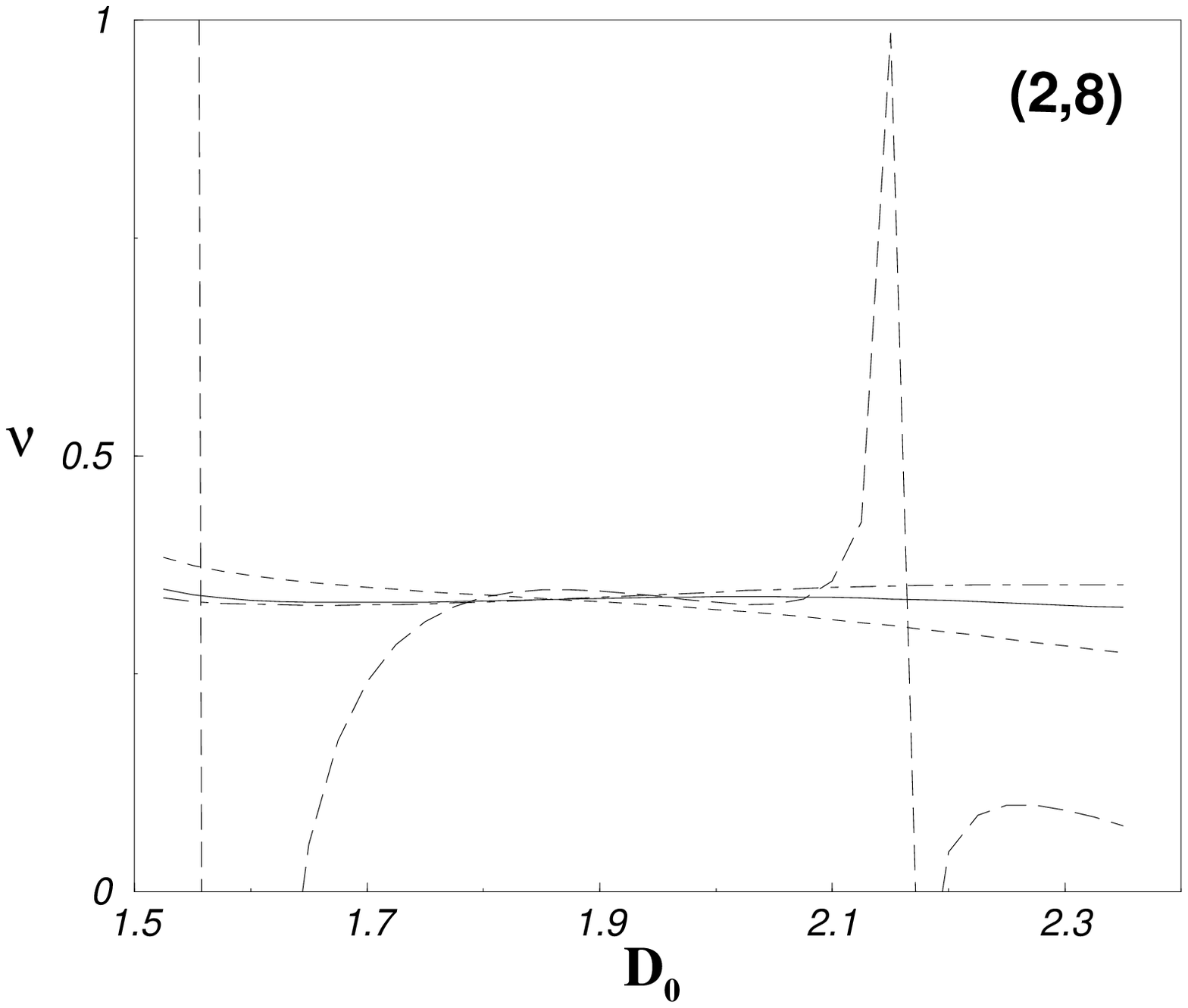}
\epsfxsize = 2.5 in \epsfbox{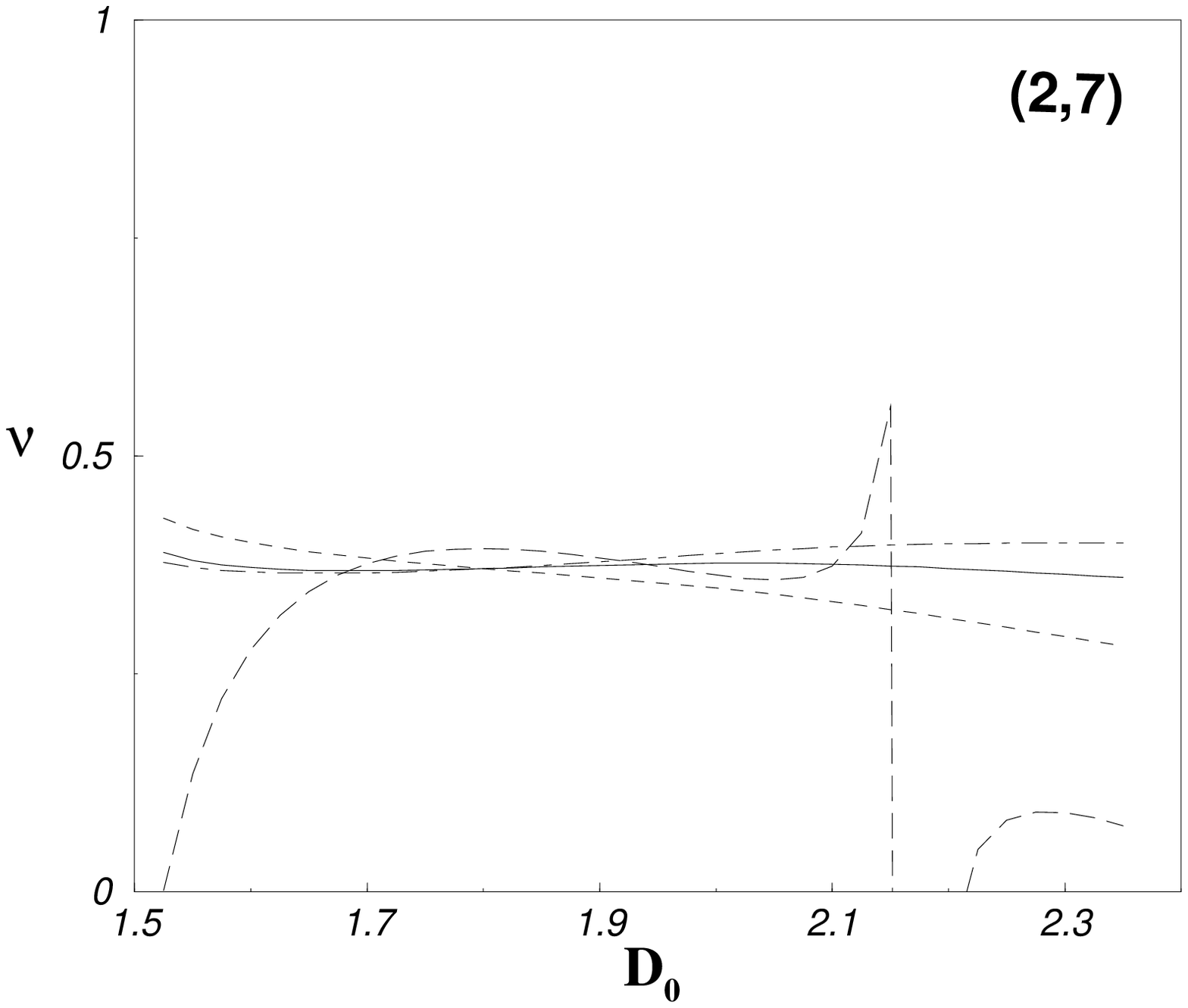}}
\centerline {\epsfxsize = 2.5in \epsfbox{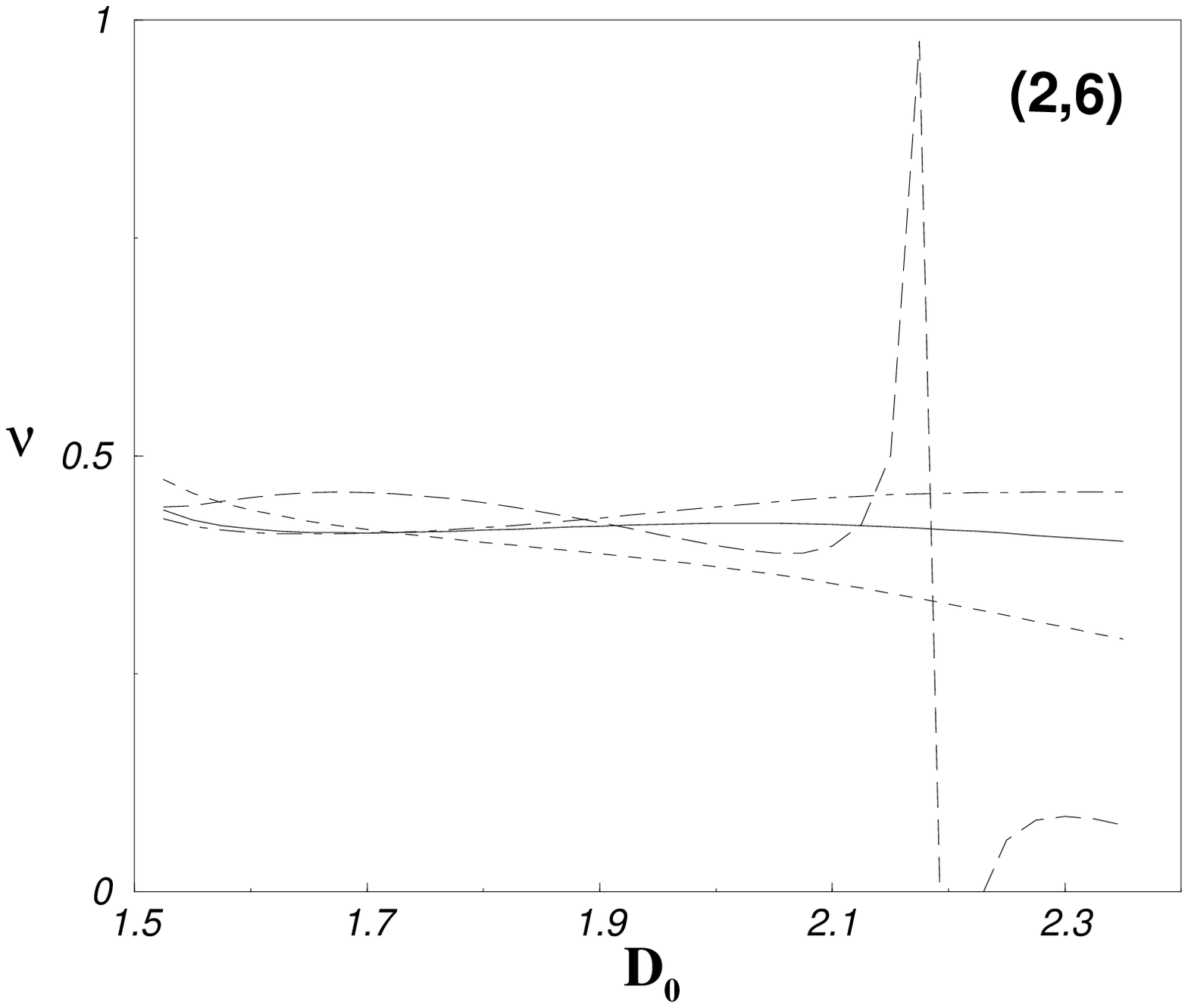}
\epsfxsize = 2.5 in \epsfbox{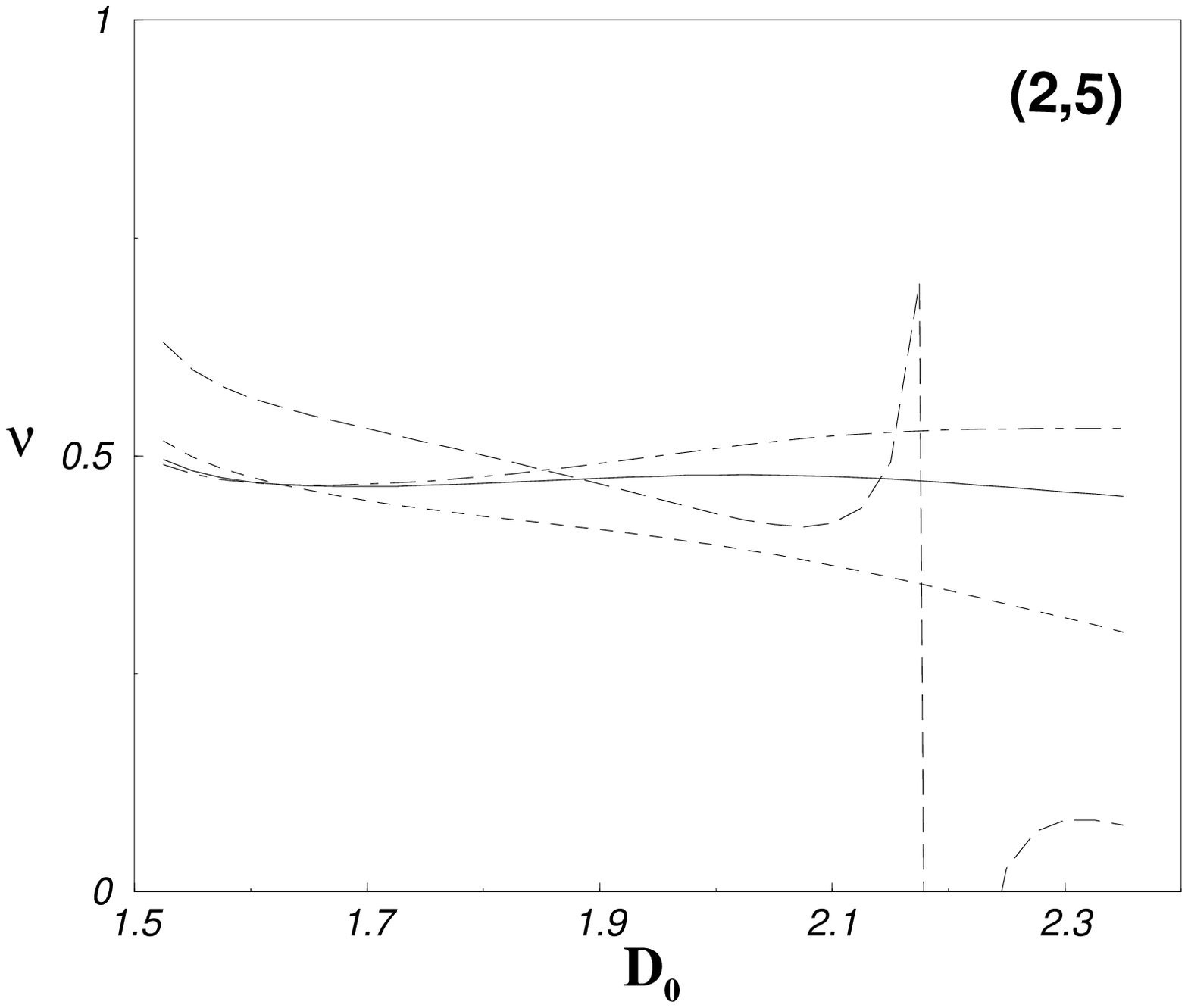}}
\centerline {\epsfxsize = 2.5in \epsfbox{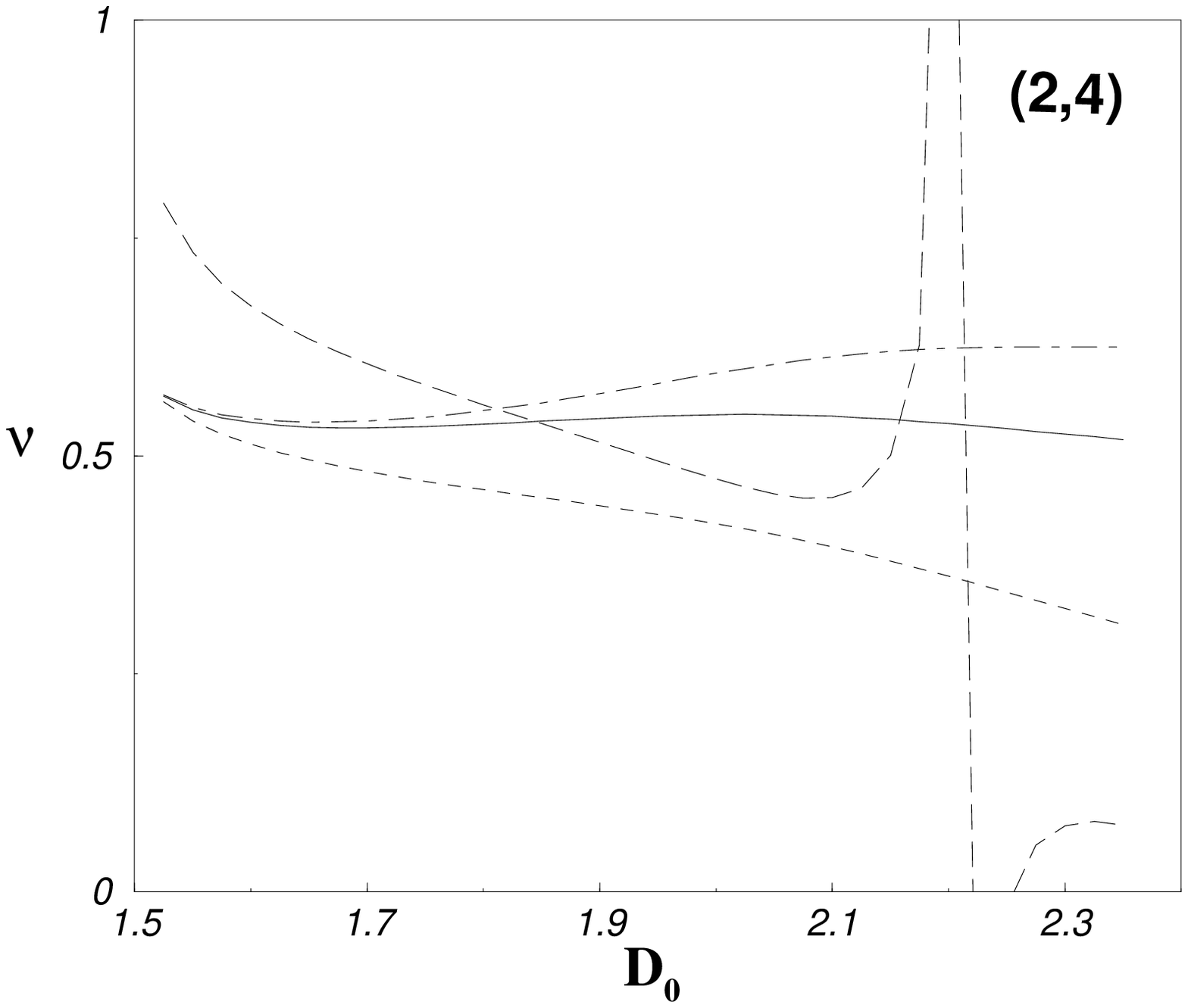}
\epsfxsize = 2.5 in \epsfbox{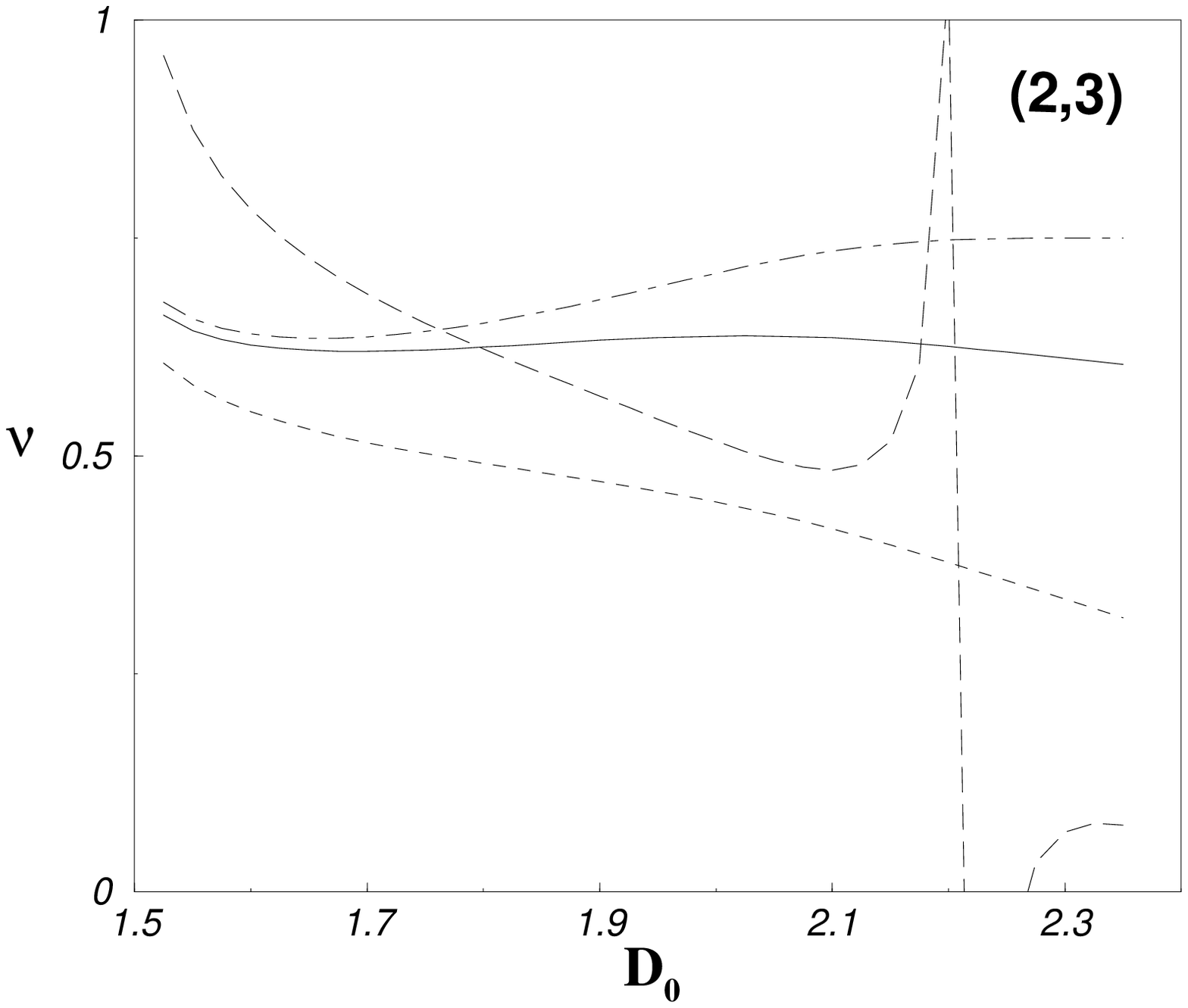}}
\caption{Calculation for the $\nu$ exponent. The Long dashed line corresponds 
to the minimal sensitivity scheme, the dashed one is A expansion, dot-dashed
corresponds to expansion C and the solid line is the D expansion.}
\label{fig__nu_extrap}
\end{figure}

The analysis leading to the extrapolation for the $\nu$ exponent,
may be summarized in
Fig.~\ref{fig__nu_extrap}. At $(D=2,d=8)$ all sets of variables A,B,C,D
give consistent results. Nevertheless, the expansion D shows the flattest 
plateau which is in complete agreement with the Flory estimate. This singles
out expansion D as the best, and in fact 
we have taken as the actual $\nu$ its value at the middle of the plateau.
Within the D expansion, results are largely independent of $D_0$. This
allows us to estimate the uncertainty in $\nu$ from its deviations from the
plateau, which is the error bar quoted in the first column
of Table~\ref{tab__EXP_fin}. Expansion C also yields compatible results
but,  as  apparent, it is not such a flat plateau. We find expansion B 
to be unreliable and the results are not even displayed. Finally, 
expansion A, almost equivalent to the naive $\vap$ expansion, shows a plateau 
coincident with D, but deviating slightly. The Hwa technique shows two 
extrema, one slightly above the Flory result, the other slightly below, so 
although the results are reasonable, we think we cannot apply it accurately
as $d$ decreases.  At $(2,7)$, the situation
for the different extrapolations is very similar;
again, expansion D gives a nice flat plateau consistent with the Flory 
estimate, and we extrapolate the best $\nu$  in the same way as in 
the $(2,8)$ case. Expansion C gives a 
result completely consistent with D, although with not such a 
flat plateau, while expansion A 
starts to deviate. The cases $(2,6)$,$(2,5)$ and $(2,4)$
follow the same trends as the previous ones, as apparent from
Fig.~\ref{fig__nu_extrap} and the results are quoted in 
Table~\ref{tab__EXP_fin}. We conclude that expansion D is a reliable
generalized $\vap $ expansion that we can confidently apply to the physical
case $(2,3)$. Our final result is quoted in Table~\ref{tab__EXP_fin}.
Let us recall that in \cite{DAVWI}
expansion D also gave the most reliable results, and it is interesting
to find the same situation here.

\begin{figure}[hp]
\centerline {\epsfxsize = 2.5in \epsfbox{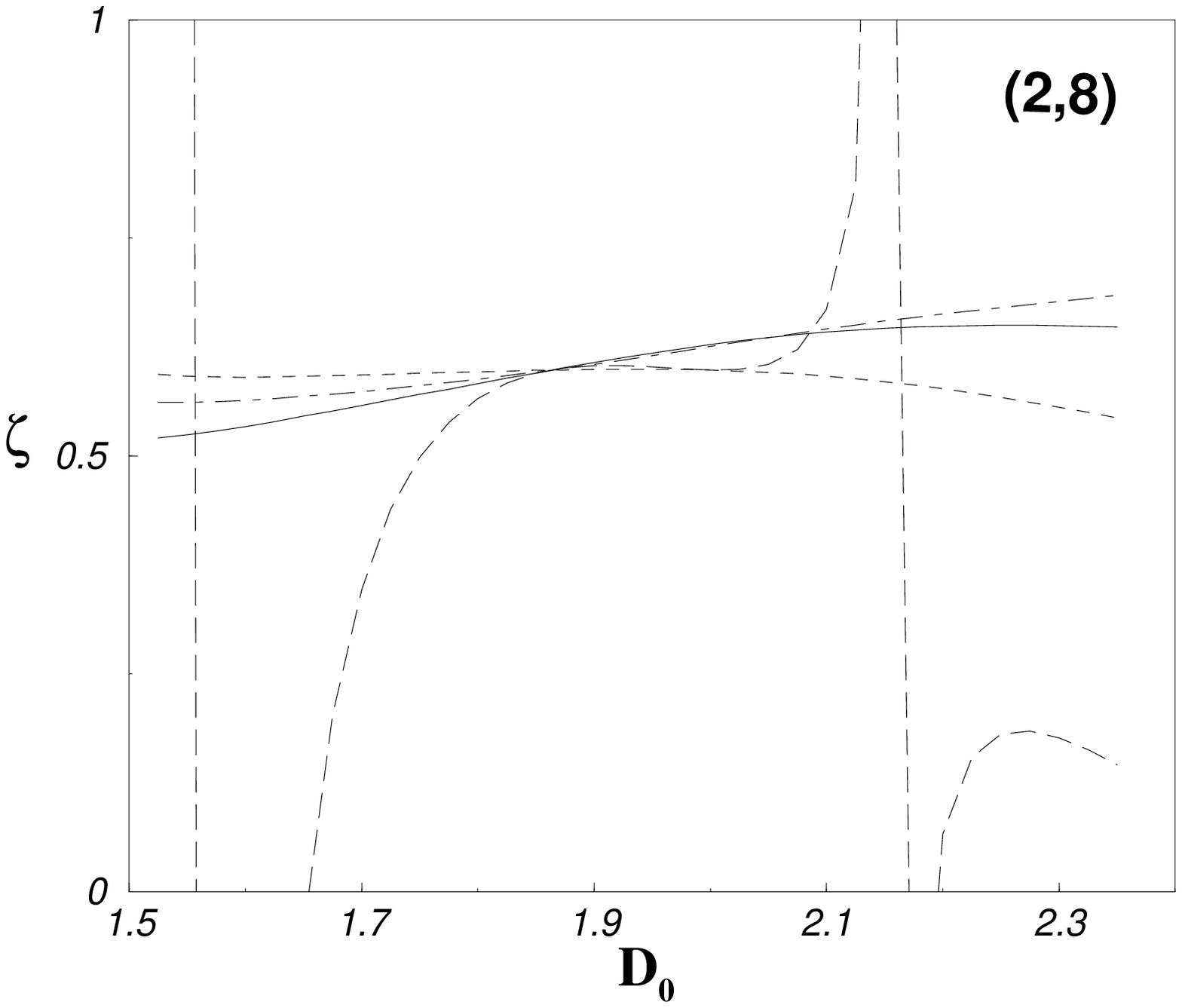}
\epsfxsize = 2.5 in \epsfbox{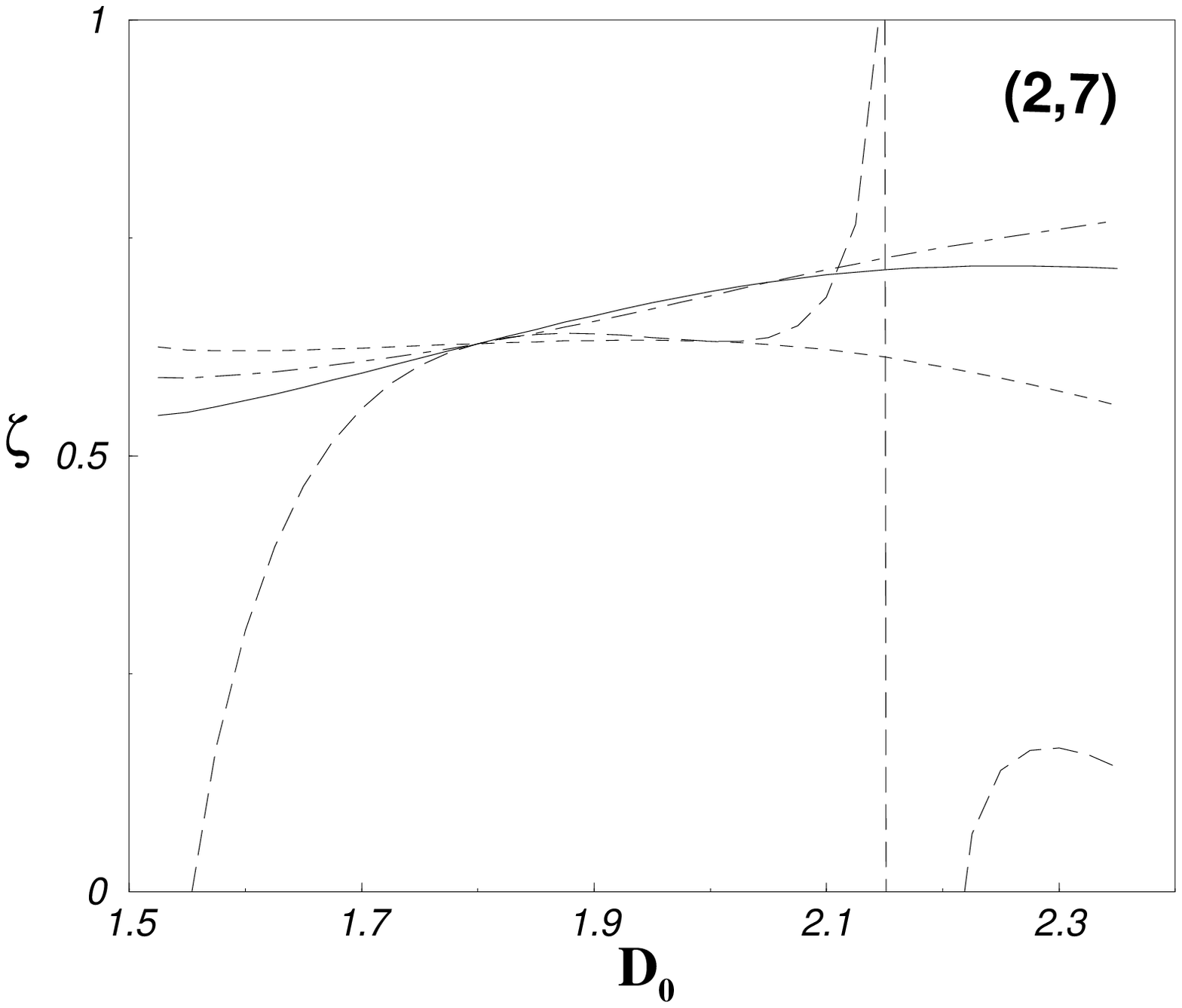}}
\centerline {\epsfxsize = 2.5in \epsfbox{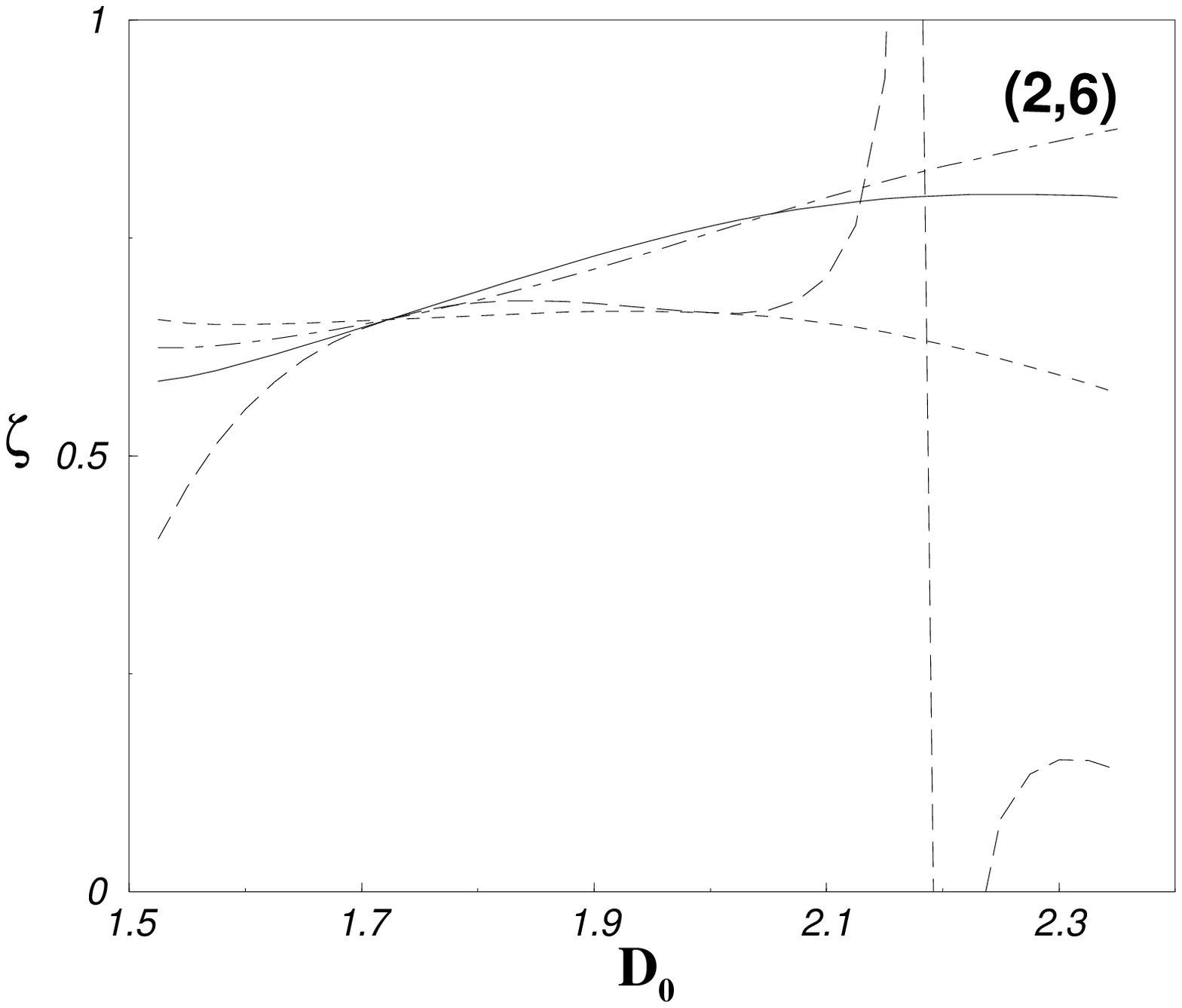}
\epsfxsize = 2.5 in \epsfbox{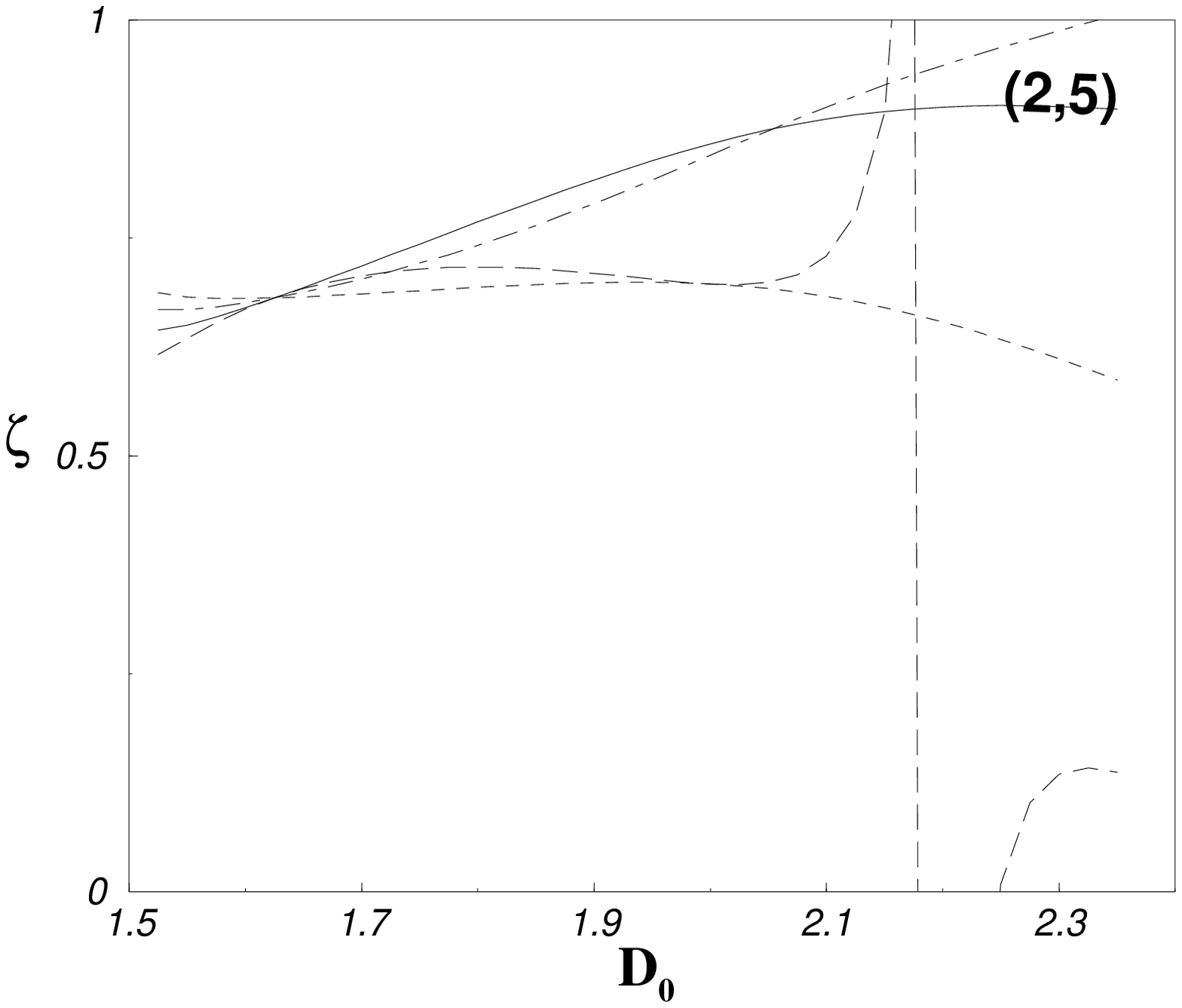}}
\centerline {\epsfxsize = 2.5in \epsfbox{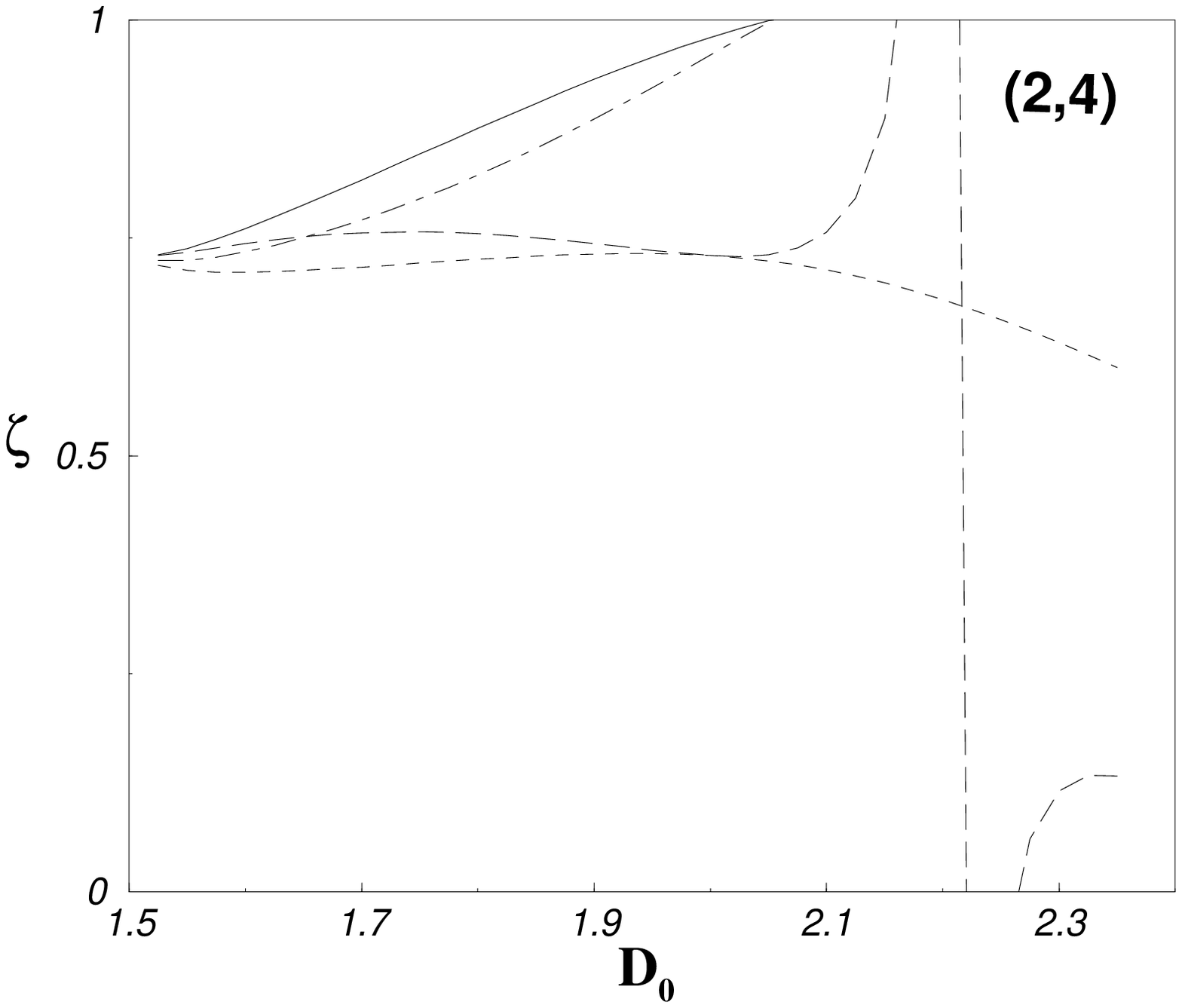}
\epsfxsize = 2.5 in \epsfbox{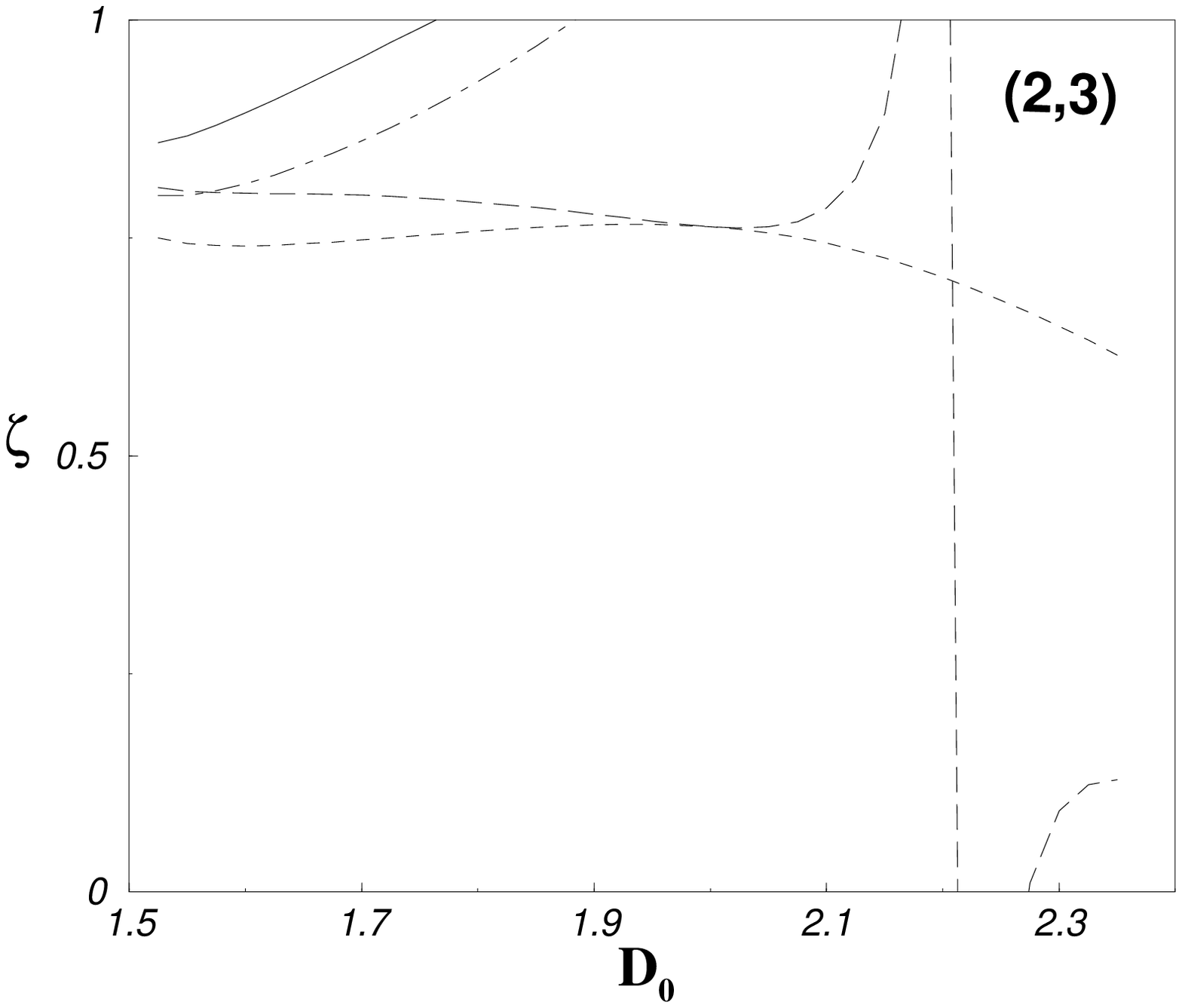}}
\caption{Calculation for the $\zeta$ exponent. The conventions are the
same as in the previous figure. }
\label{fig__zeta_extrap}
\end{figure}

Concerning the $\zeta$ exponent, the situation is different. We
find the best results applying Hwa's technique. For small $d$ 
any of the estimates A,B,C,D exhibits a large enough plateau, and
consequently, we are not confident enough of their robustness. The estimates
quoted in Table~\ref{tab__EXP_fin} are those obtained from the Hwa method. As
shown in Fig.~\ref{fig__zeta_extrap} we find two points where
$\frac{\partial{\nu}}{\partial D_0}=0$, one for $D_0<2$ the
other for $D_0>2$. Our actual estimate corresponds to the case $D_0<2$
since, on one hand it agrees slightly better with the Flory estimate
for large $d$, and on the other the curve $\vap=0$ seems intuitively closer 
to the actual point. 

Finally, it is reassuring that the values we obtain for $\nu$ and $\zeta$,
although computed using different extrapolations, are compatible with
the scaling relations Eq.~\ref{all_scaling}.

\begin{table}[htb]
\centerline{
\begin{tabular}{|c||l|l||l|l|}
\multicolumn{1}{c}{$d$}   & 
\multicolumn{1}{c}{$\nu$}   & \multicolumn{1}{c}{$\nu_{Flory}$} & 
\multicolumn{1}{c}{$\zeta$} & \multicolumn{1}{c}{$\zeta_{Flory}$} \\\hline
8  & $0.333(5) $ & $0.333$ & $0.60$  & $0.600$  \\\hline
7  & $0.374(8) $ & $0.375$ & $0.64$  & $0.643$  \\\hline
6  & $0.42(1)  $ & $0.429$ & $0.68$  & $0.692$  \\\hline
5  & $0.47(1)  $ & $0.500$ & $0.72$  & $0.750$  \\\hline
4  & $0.54(2)  $ & $0.600$ & $0.76$  & $0.818$  \\\hline
3  & $0.62(2)  $ & $0.750$ & $0.80$  & $0.900$  \\\hline
\end{tabular}}
\caption{Final results for critical exponents.}
\label{tab__EXP_fin}
\end{table}

\subsubsection{Corrections to the Flory estimate}

The Flory approximation, which certainly works well for polymers, also
provides valuable insight in the case of membranes. The basic
approximation assumes that elastic energies are comparable to
self-avoiding energies. For self-consistency we then require that
$Z_{\perp}=Z_b$.
This extra condition fixes $\delta$ and $\nu$ to be
\bea\label{google}
\delta_F&=&\frac{-4 \vap}{4 + (D-1)(d+3)} 
\nonumber\\
\nu_{Flory}&=&\frac{D+1}{d+1} \ .
\eea 
It is also interesting to analyze the corrections to the Flory result.
We can write at the fixed point the following equivalent definition
for $\delta$,
\be\label{delta_flory}
\delta=\delta_F-\frac{4}{4+(D-1)(d+3)} \mu \frac{d}{d \mu} 
\log(\frac{Z_b}{Z_{\perp}}) \ .
\ee
Expanding to first order in $\vap$ it follows that  
\be\label{nu_flory}
\nu_F(D)=\frac{5-2D}{4}+\frac{17-4D}{3(D+1)}\nu_1(D)\vap ,
\ee
where $\nu_1(D)$ is defined in Eq.~\ref{nu_loop}. Coincidentally the 
extra factor appearing on the r.h.s of the previous equation is just $1$ at
$D=2$, so Eq.~\ref{nu_loop} and Eq.~\ref{nu_flory} give the same 
result at $D=2$, a result already noticed in \cite{BG}.

\begin{figure}[hp]
\centerline {\epsfxsize = 2.5in \epsfbox{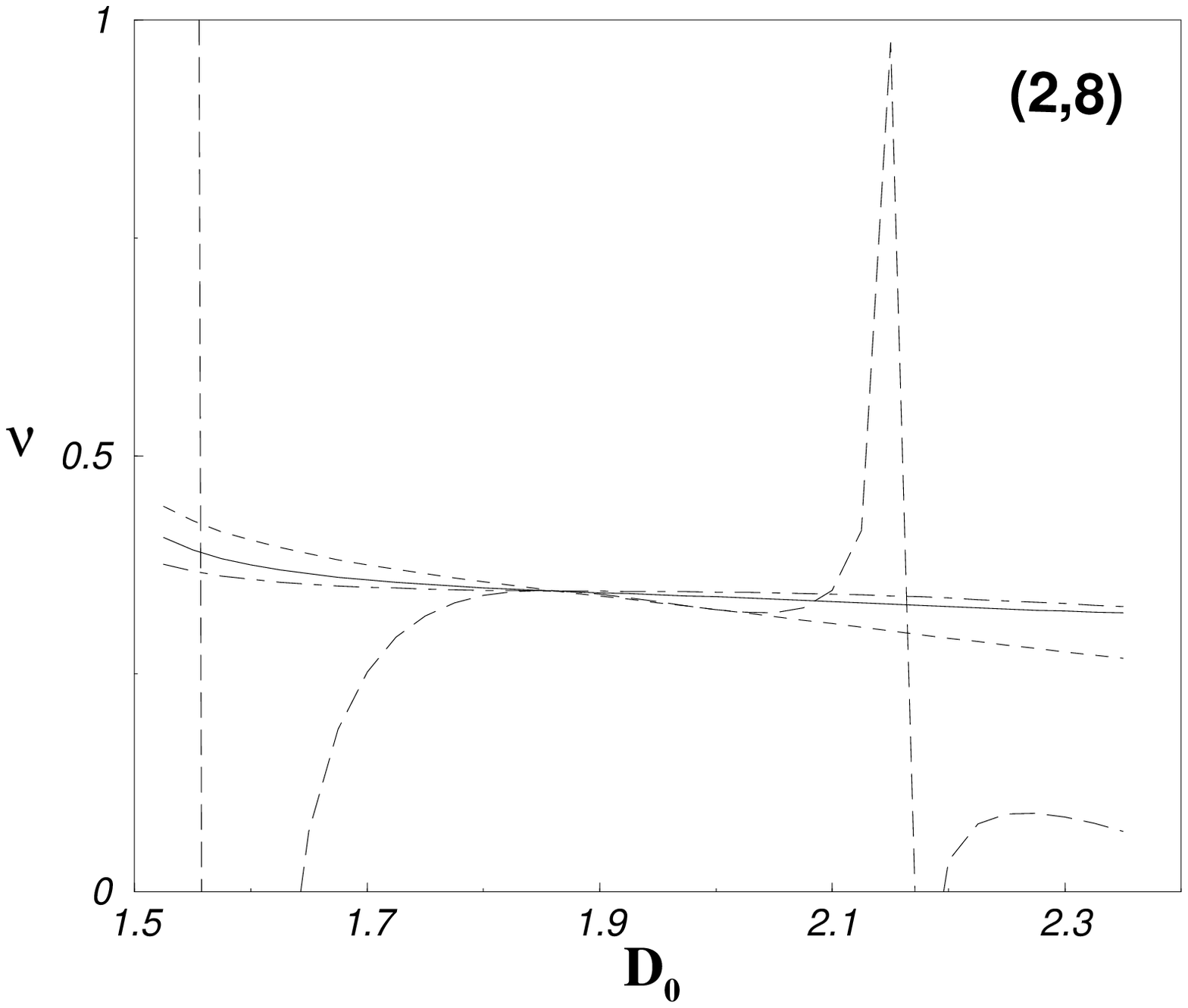}
\epsfxsize = 2.5 in \epsfbox{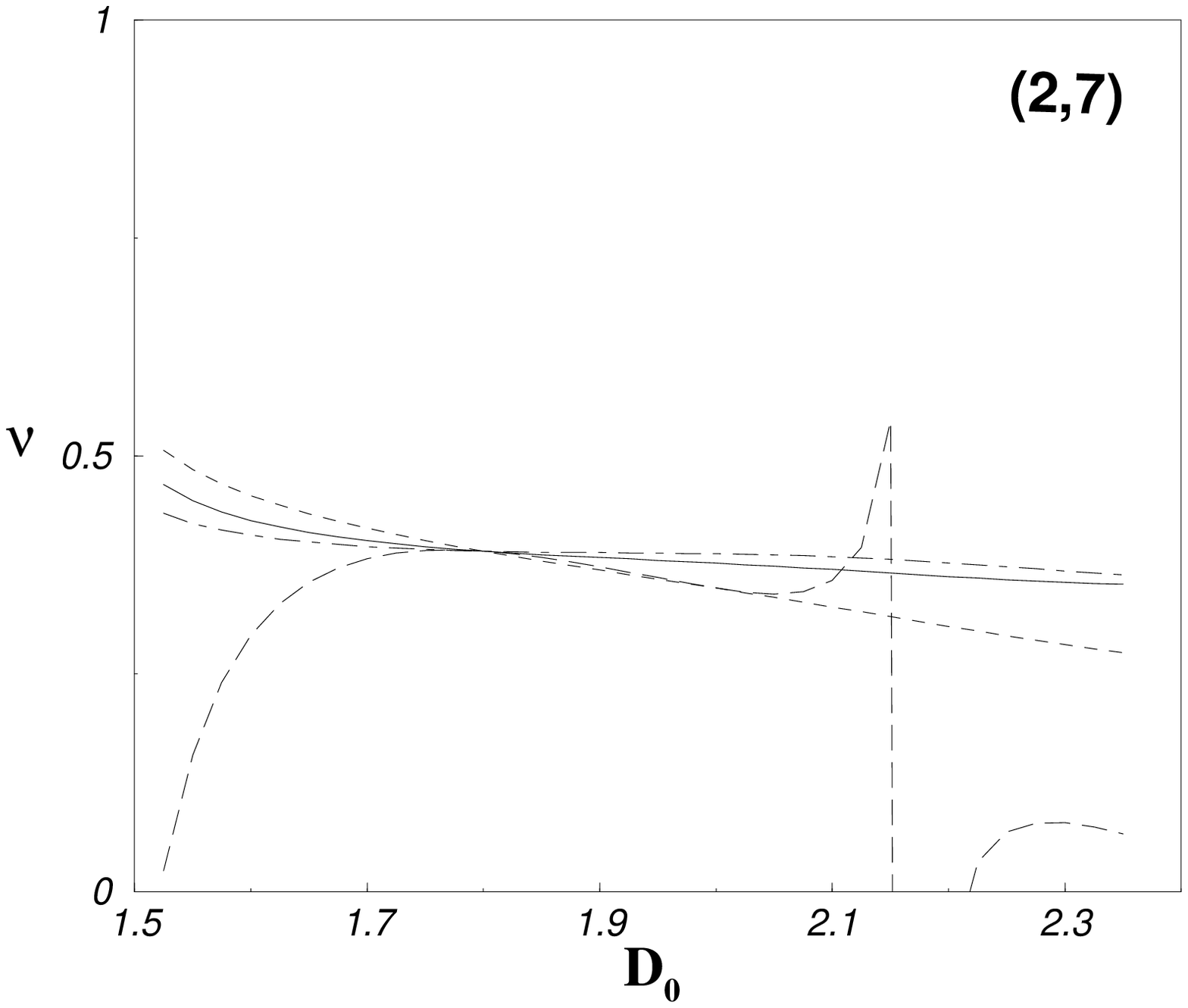}}
\centerline {\epsfxsize = 2.5in \epsfbox{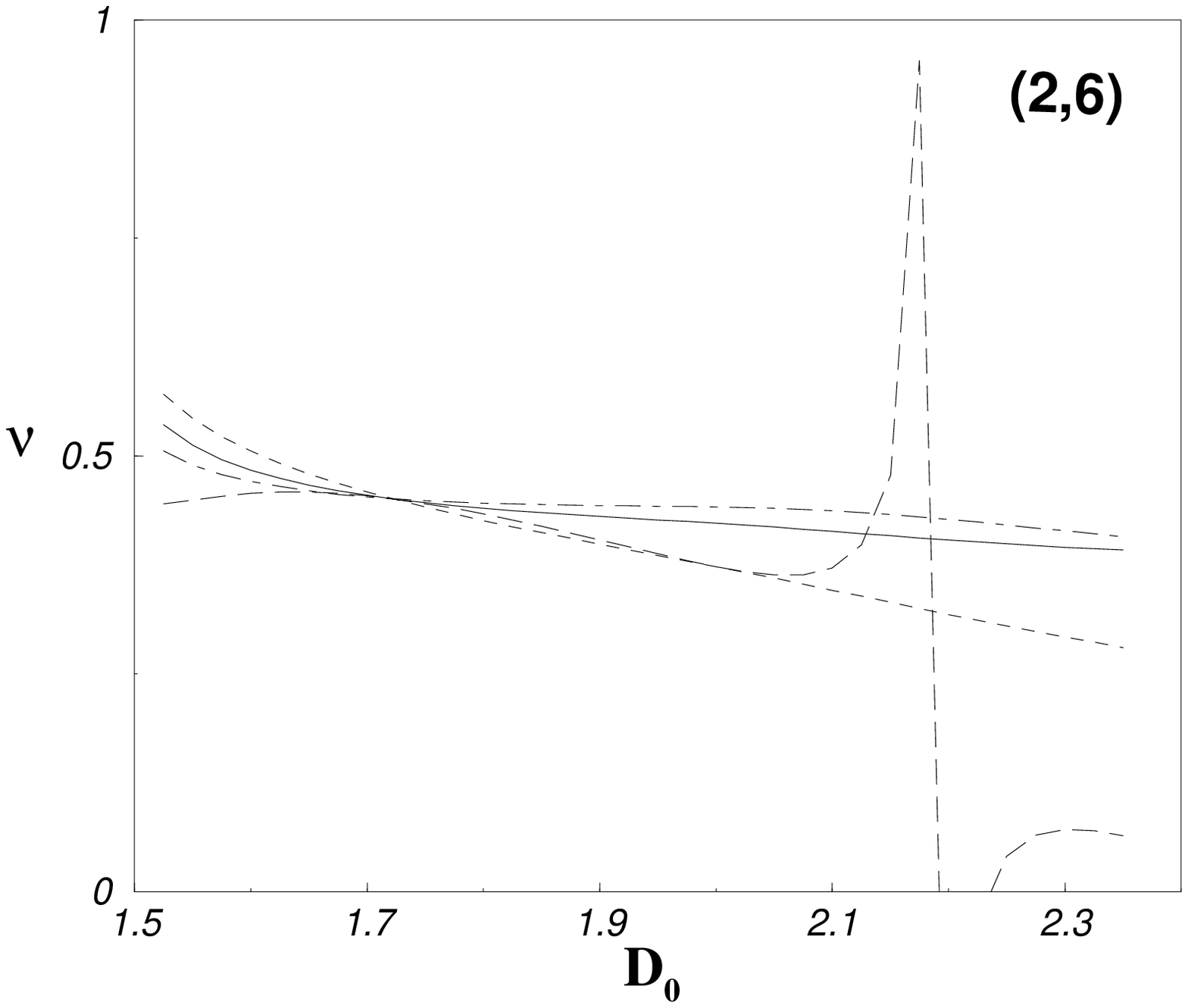}
\epsfxsize = 2.5 in \epsfbox{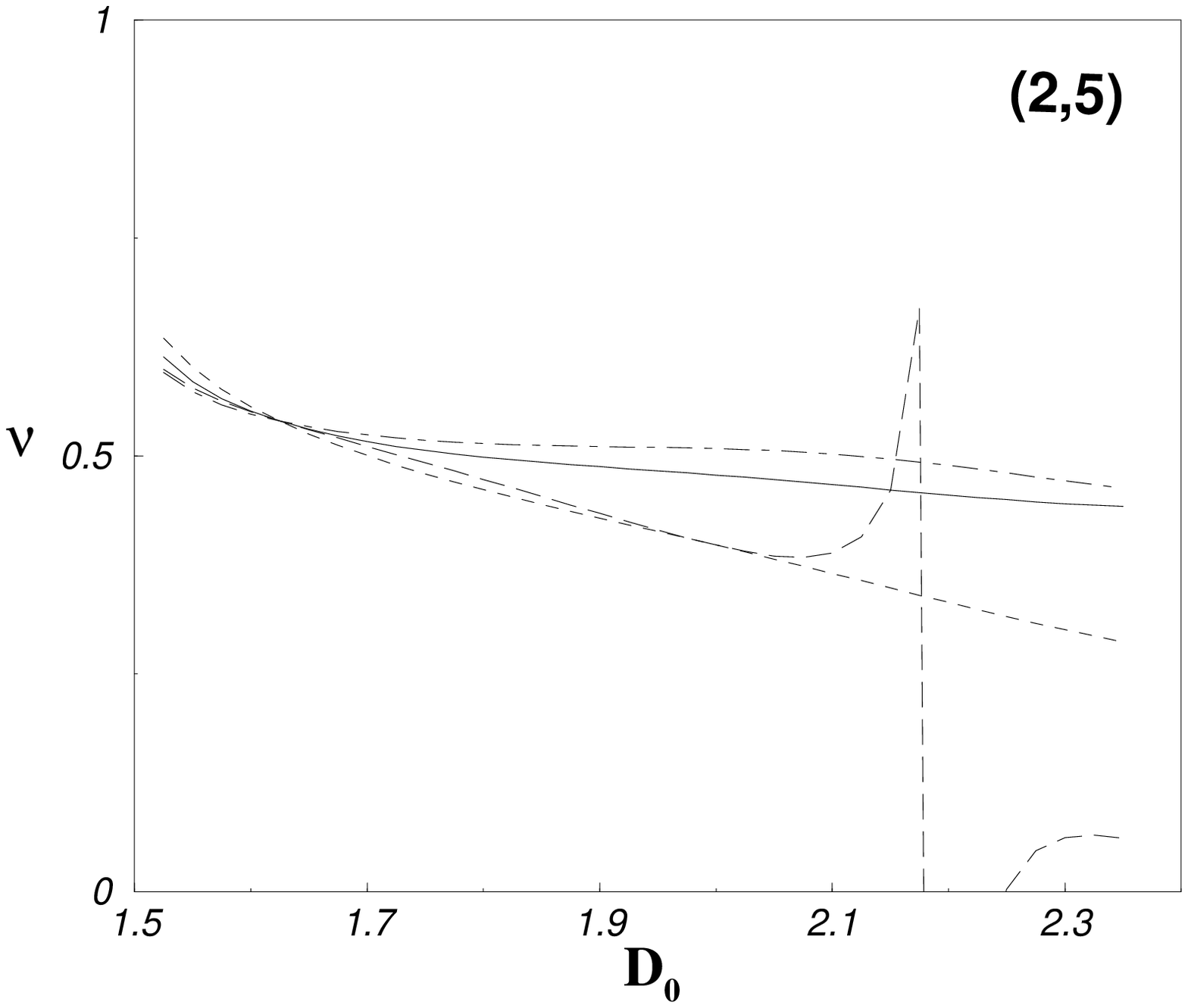}}
\centerline {\epsfxsize = 2.5in \epsfbox{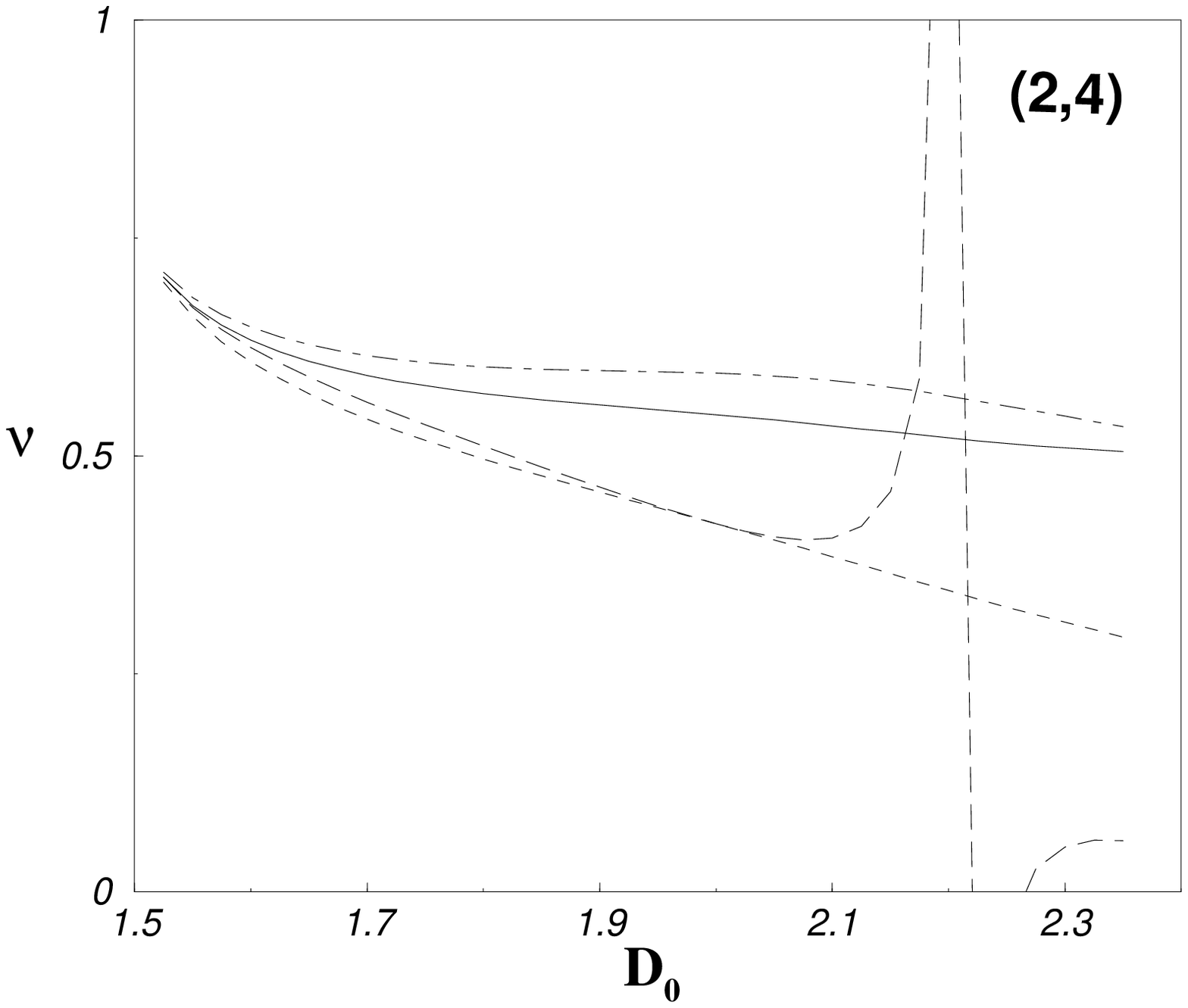}
\epsfxsize = 2.5 in \epsfbox{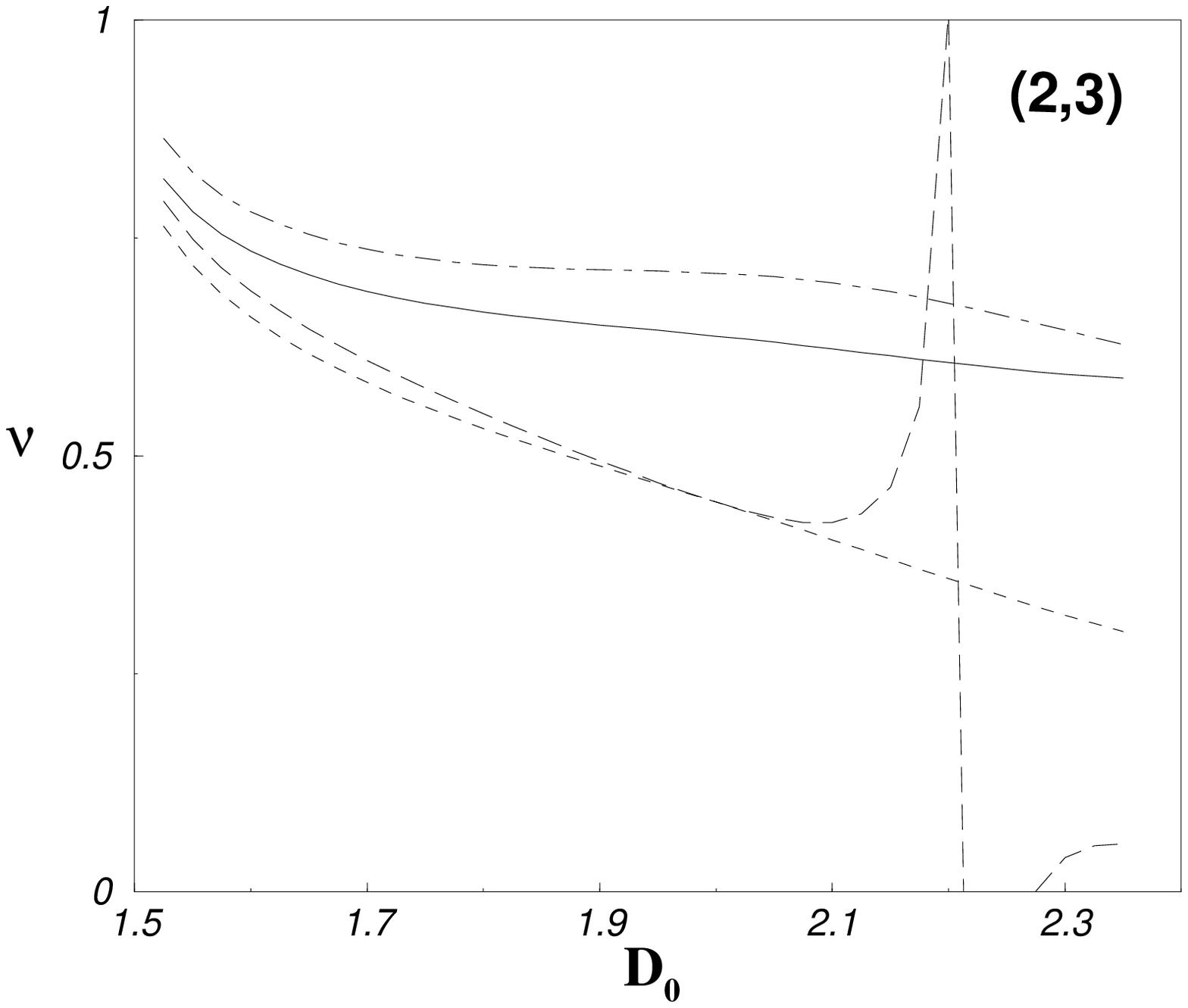}}
\caption{Corrections to the Flory estimate for $\nu$.}
\label{fig__nu_flory}
\end{figure}

We can apply the usual machinery to extract critical exponents. 
We start by examining the $(2,8)$ case. From Fig.~\ref{fig__nu_flory} 
it is clear that all expansions give compatible results,
although expansion C produces the flattest plateau.
We use this expansion to extract our best $\nu$ within this approximation.
Using the fluctuations in the plateau, as a function of $D_0$, to
estimate errors, we realize that results in this case are not as accurate, 
although still compatible
with previous estimates and with its Flory value. From analyzing this 
case we see also that this expansion tends to give a slight overestimate. 
The same situation holds as $d$ decreases, as shown in
Table~\ref{tab__EXP_comp}. 
Overall, results remain close to the Flory estimate although with a
larger uncertainty. They are still compatible with
the values quoted in Table~\ref{tab__EXP_fin}, which we regard as our most 
accurate determinations.

\begin{table}[htb]
\centerline{
\begin{tabular}{|c||l|l|l|l|l|}
\multicolumn{1}{c}{$d$}     & 
\multicolumn{1}{c}{$\nu$}   & \multicolumn{1}{c}{$\nu_F$ } & 
\multicolumn{1}{c}{$\nu_V$} & \multicolumn{1}{c}{$\nu_{Flory}$} \\\hline
8  & $0.333(5)  $ & $0.34(1)$  & $0.34(1)$  & $0.333$ \\\hline
7  & $0.374(8)  $ & $0.39(2)$  & $0.39(2)$  & $0.375$ \\\hline
6  & $0.42(1)   $ & $0.44(2)$  & $0.44(4)$  & $0.429$ \\\hline
5  & $0.47(1)   $ & $0.51(3)$  & $0.51(5)$  & $0.500$ \\\hline
4  & $0.54(2)   $ & $0.60(4)$  & $0.60(6)$  & $0.600$ \\\hline
3  & $0.62(2)   $ & $0.71(6)$  & $0.70(9)$  & $0.750$ \\\hline
\end{tabular}}
\caption{Comparison of the different extrapolations for $\nu$}
\label{tab__EXP_comp}
\end{table}

\subsubsection{Corrections to the Gaussian variational approximation}

Another approach that has relatively successful in dealing with
problems with self-avoidance is the Gaussian variational approximation
\cite{Gauss}.
It consists in approximating the exact density functional by the best
possible quadratic weight for the field $\vec h$. It amounts to 
assuming that the self-avoiding term is not renormalized, that is
$Z_b=1$. The quantity $\delta$ and the gyration radius exponent within
this approximation were first computed in \cite{RT1} with result
\bea\label{winkel}
\delta_V&=&\frac{-4 \vap}{(D-1)(d+3)} 
\nonumber\\
\nu_{var}&=&\frac{7(D-1)}{(3d-5)} \ .
\eea 
The value for the physical tubule is $\nu_{var}=\frac{7}{4}$. This is 
clearly unphysical, being larger than one, but the accuracy of the
Gaussian variational approximation should improve for large $d$, since
it is essentially a large{-}$d$ expansion.
As with the Flory approximation, we may determine the corrections
to the Gaussian variational approximation within the $\vap${--}expansion.
From
\be\label{delta_variat}
\delta=\delta_V-\frac{4}{(D-1)(d+3)} \mu \frac{d}{d \mu} 
\log(Z_b) \ ,
\ee
which consistently at lowest order in $\vap$ results in
\be\label{nu_variat}
\nu_V(D)=\frac{5-2D}{4}+\frac{\nu_1(D)}{D-1}\vap ,
\ee
where $\nu_1(D)$ is defined in Eq.~\ref{nu_loop} and again
the extra factor appearing on the r.h.s is just $1$ at
$D=2$, so Eq.~\ref{nu_loop} and Eq.~\ref{nu_variat} are equal  
at $D=2$, as reported in \cite{BG}.

\begin{figure}[hp]
\centerline {\epsfxsize = 2.5in \epsfbox{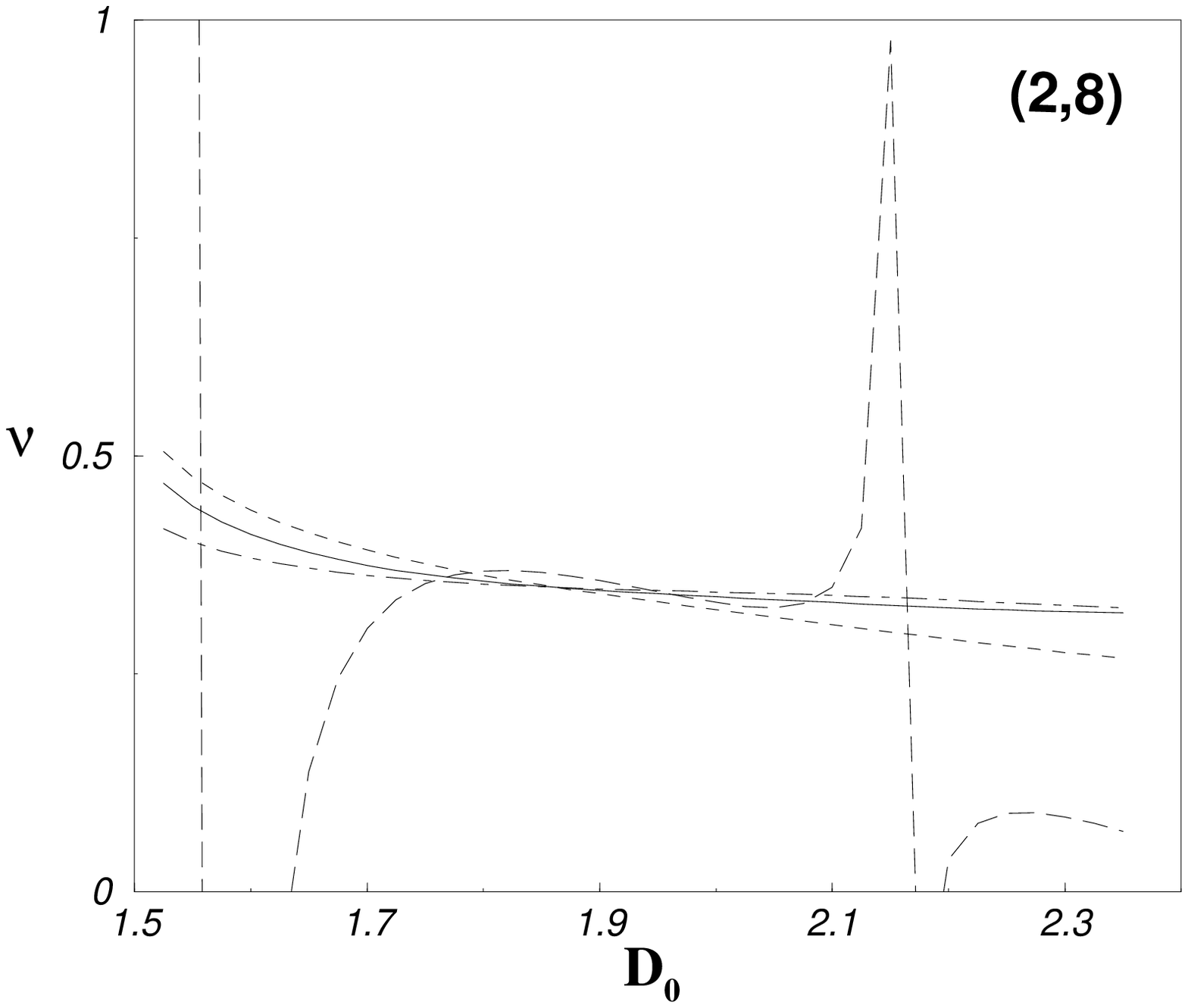}
\epsfxsize = 2.5 in \epsfbox{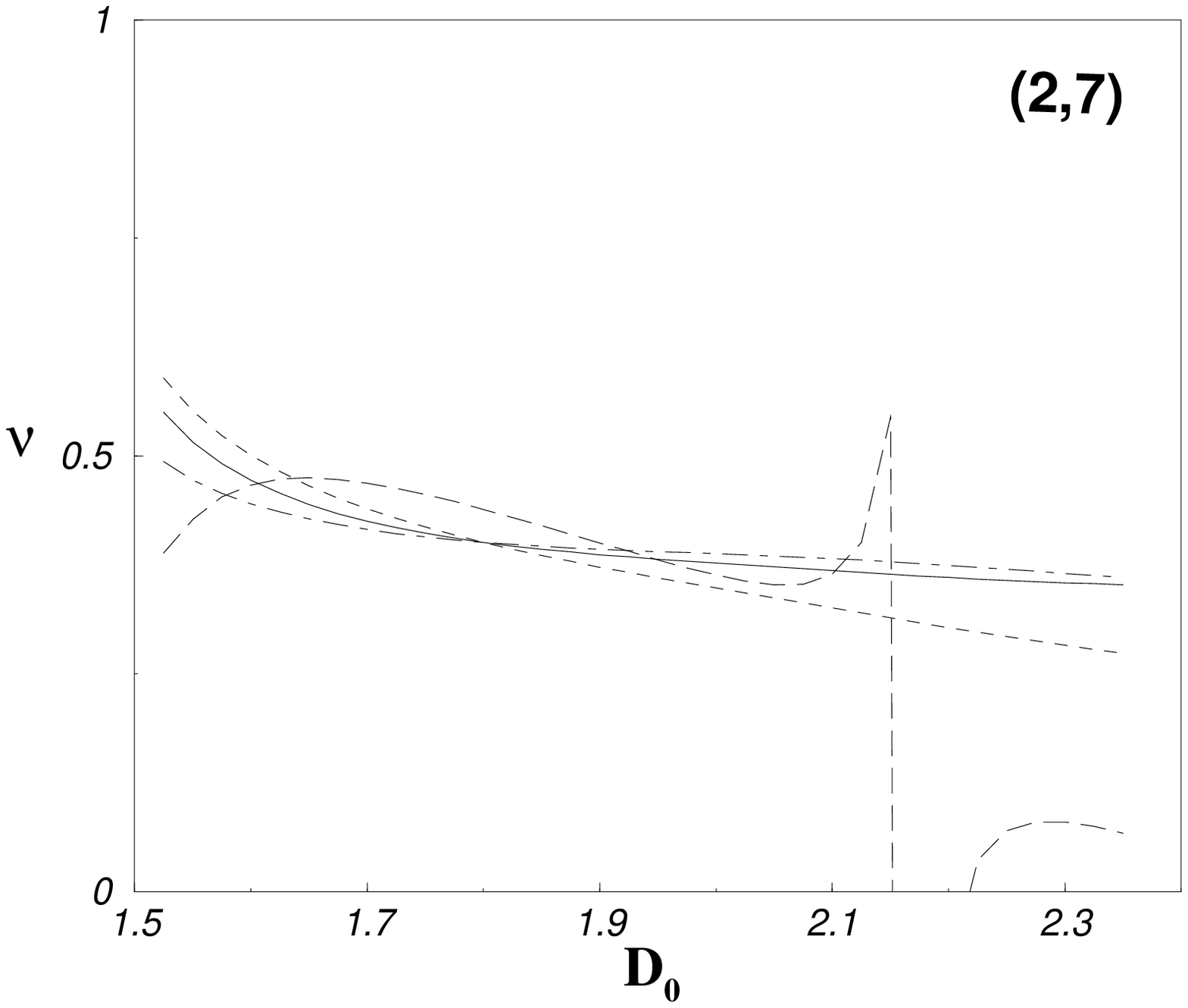}}
\centerline {\epsfxsize = 2.5in \epsfbox{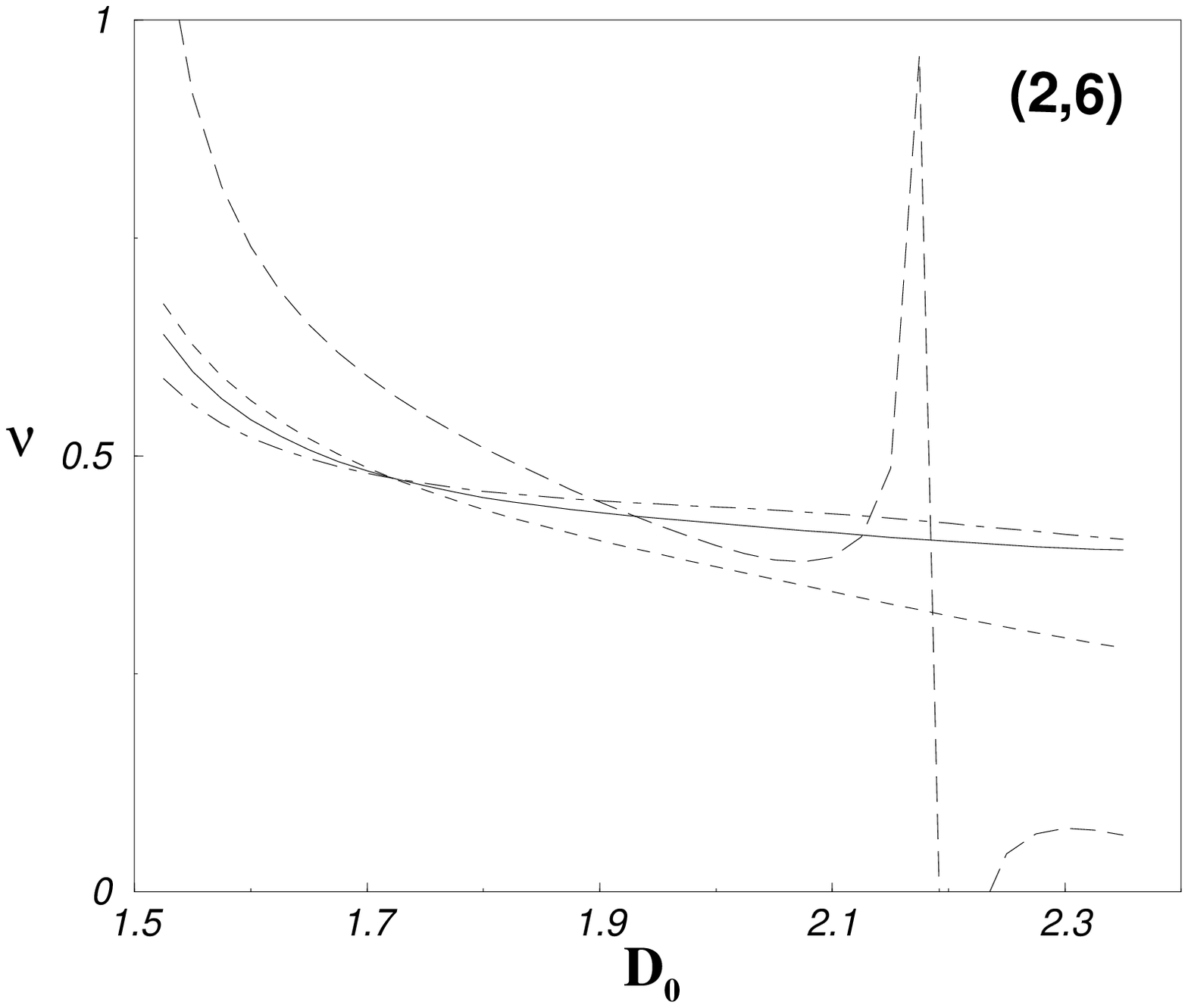}
\epsfxsize = 2.5 in \epsfbox{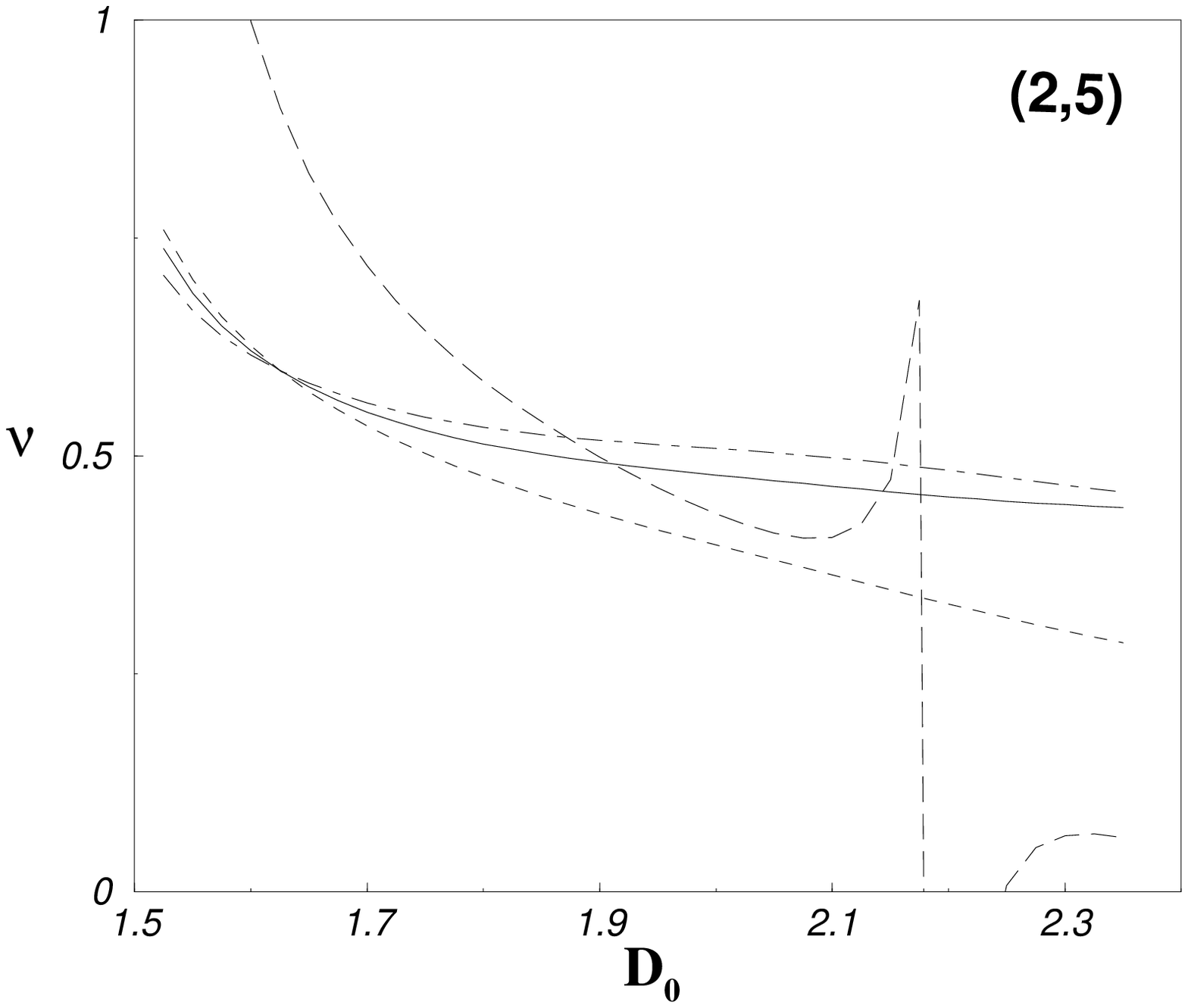}}
\centerline {\epsfxsize = 2.5in \epsfbox{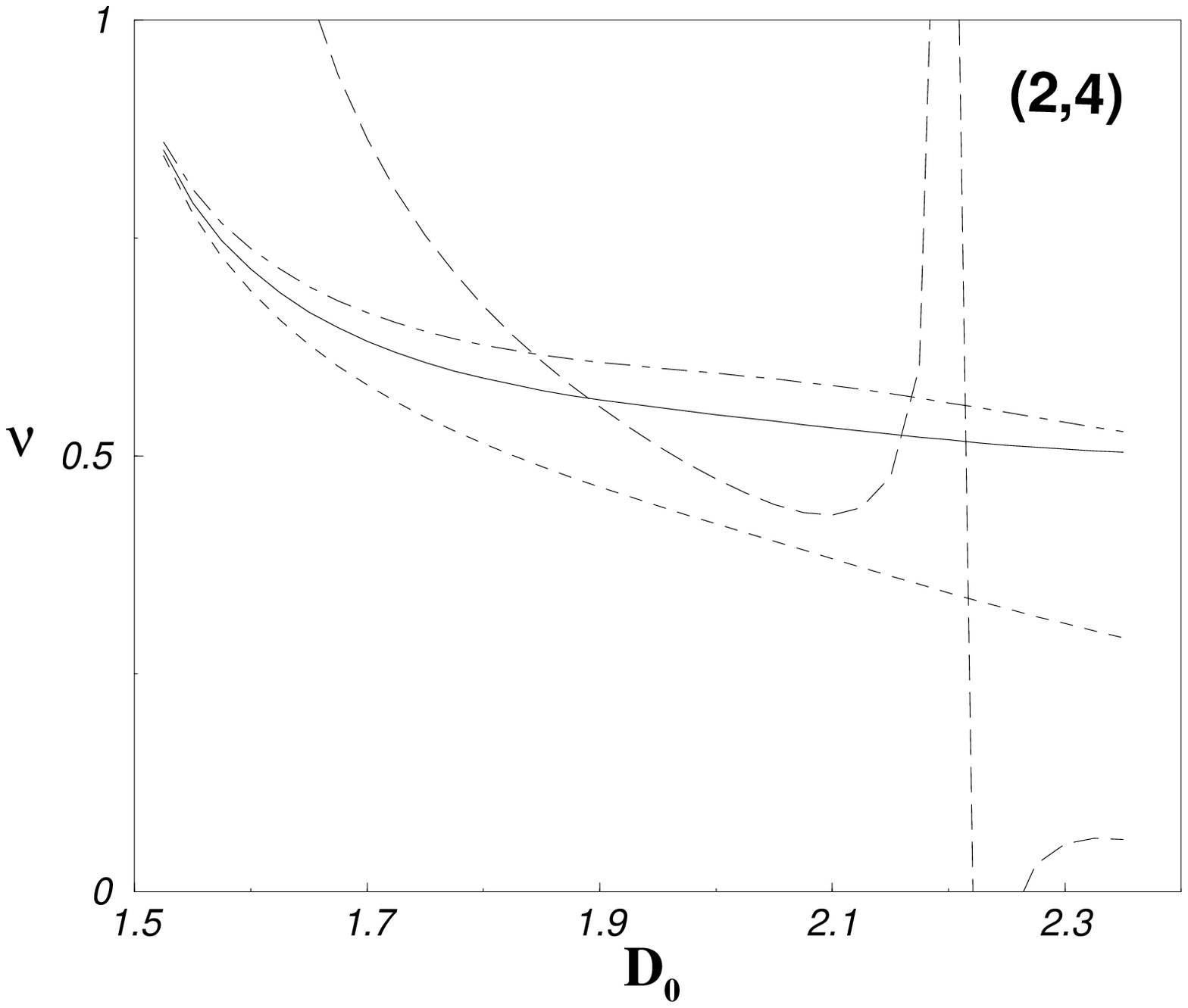}
\epsfxsize = 2.5 in \epsfbox{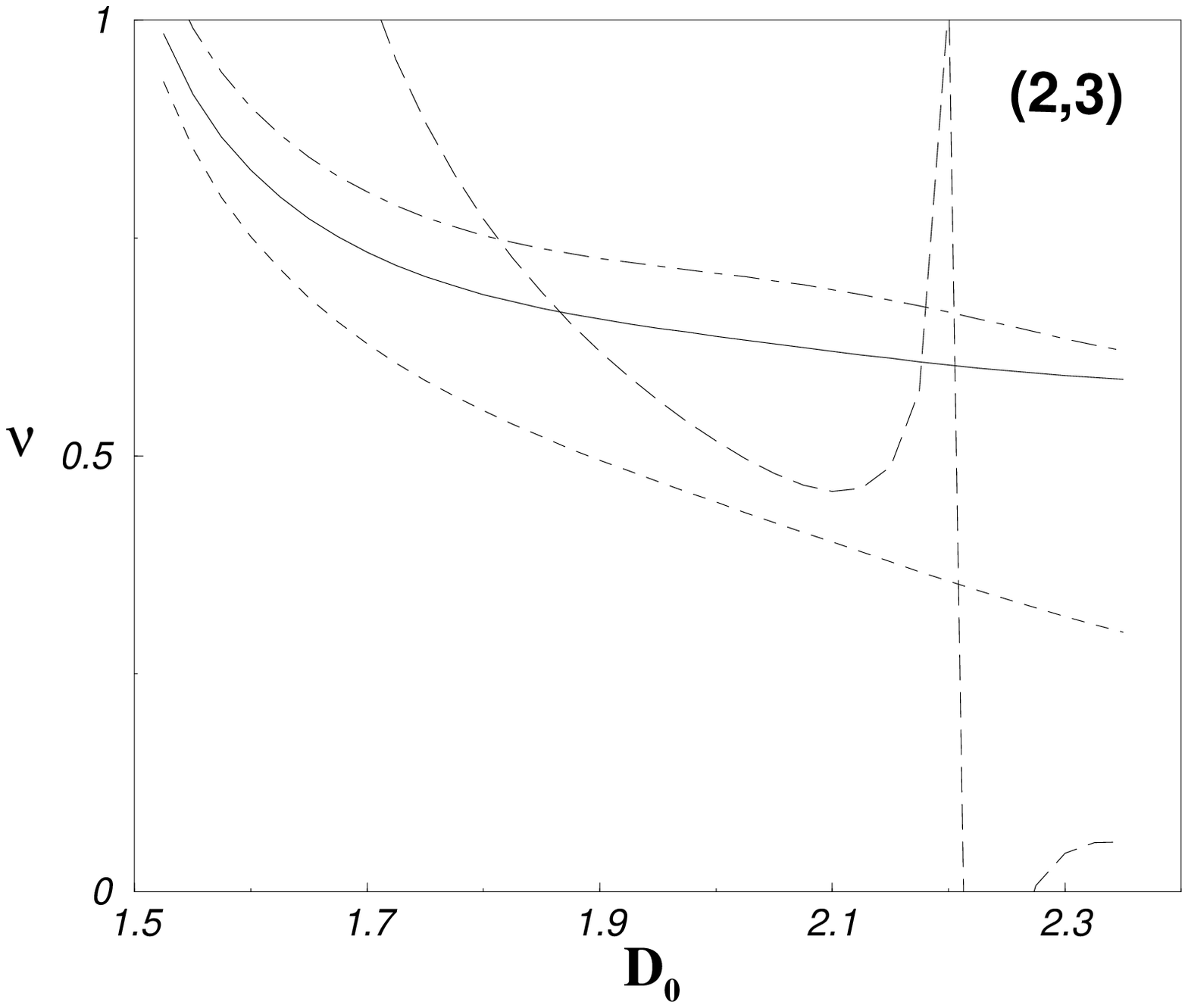}}
\caption{Corrections to the variational estimate for $\nu$.}
\label{fig__nu_variat}
\end{figure}

Our extrapolations are summarized in Fig.~\ref{fig__nu_variat}.
The Gaussian variational approximation turns out to be
the least accurate of our determinations. At $(2,8)$ all 
different extrapolations deliver equivalent results, but the expansion
C shows the flattest plateau. The situation is the same for the cases 
$(2,7)$,$(2,6)$ and $(2,5)$, but at $(2,4)$ and $(2,3)$, there are
sizeable variations and there is no a clear plateau. For the sake of 
completeness, we extrapolate our results from this pseudo-plateau
where variations are small, and quote an error bar from its variations,
which are the results quoted in Table~\ref{tab__EXP_comp}.

\section{Conclusions}\label{SECT__Con}

In this paper our first task was to identify the simplest free energy 
containing all the relevant operators controlling the large-distance
physics of the tubular phase of anisotropic membranes. In this
analysis essential use was made of rotational symmetries. 
Although our analysis may be modified by the existence of more complicated phase
diagrams with non-perturbative fixed points in the spirit of
\cite{RT2}, we believe that the model treated here reveals essential
features of the physics of the anisotropic tubular phase.

Finally we completely characterized the phase diagram
and calculated the critical exponents by generalizing the
$\vap${--}expansion introduced in \cite{BG}.
For the physical self-avoiding tubule we find 
\bea\label{ultimate}
\nu&=&0.62
\\
\zeta&=&0.80
\\
z&=&0.75
\\
\zeta_{u}&=&-0.33
\\
\eta_u&=&1.65
\\
\eta_{\perp}&=&1.0
\eea
Further improvement would necessitate
a two-loop calculation for arbitrary $D$ and would provide a valuable
check of our extrapolation.

These predictions may be tested via an extension of the numerical
simulations described in \cite{BFT} to the much more demanding model
with self-avoidance. These simulations are currently in progress.
We hope the concreteness of the calculations presented here will
inspire further work in the rich field of the physics of anisotropic extended
manifolds.  

\vspace{1cm}
\centerline{\large ACKNOWLEDGEMENTS}
\vspace{0.75cm}
We thank Emmanuel Guitter, Leo Radzihovsky and John Toner for several
valuable discussions on the subject matter of this paper. 
The research of M.B and A.T was supported by the U.S. Department of Energy under 
Contract No. DE-FG02-85ER40237. 

\appendix
\section{Appendix}\label{SECT__App}

\begin{figure}[htb]
\epsfxsize=5in\centerline{\epsfbox{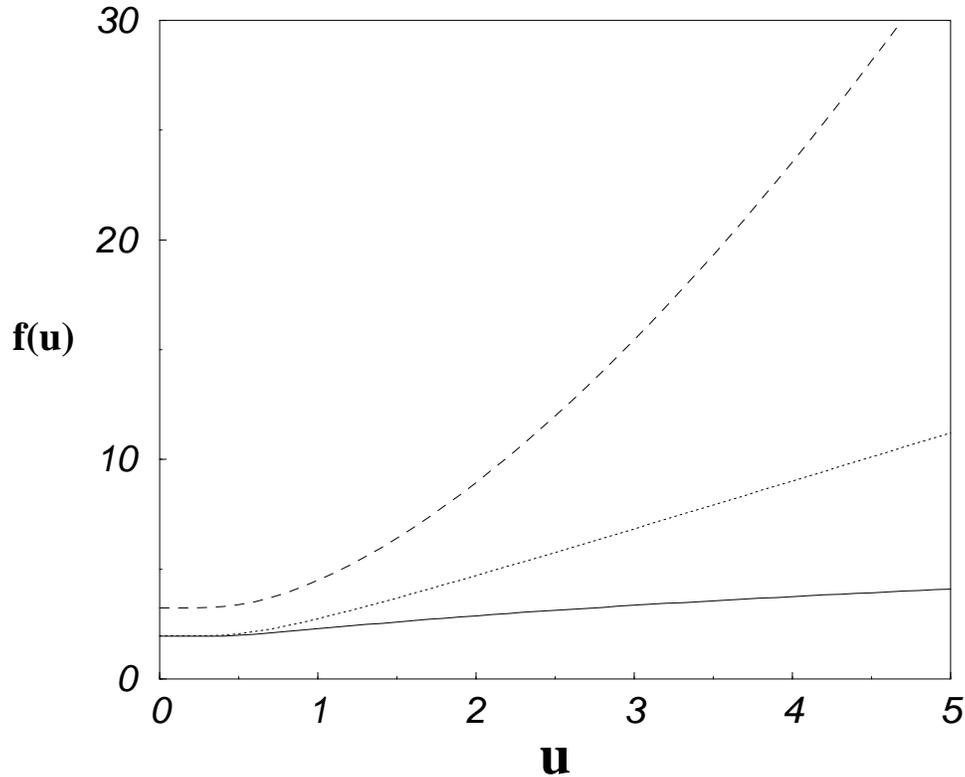}}
\caption{$f(u)$ for different values of $D$. The dashed line is
$D=1.7$, the dotted line $D=2.0$ and the solid line $D=2.3$.}
\label{fig__fu}
\end{figure}

In this appendix we discuss the analytical properties of several 
functions that arise in the evaluation of the quantity $I(D)$.
We follow closely the methods of \cite{BG}, and rewrite Eq.~\ref{def_ID} 
as
\be\label{new_def_ID}
I(D)=\frac{1}{\Gamma(\frac{4D-3}{5-2D})}\int^{+\infty}_{0} dz F(z)^2
\ ,
\ee
with
\be\label{def_FZ}
F(z)=z^{\frac{3D-4}{5-2D}}\int^{\infty}_0 du u^{2D-3} e^{-zf(u)} \ ,
\label{def_fu}
\ee
and 
\bea\label{def_fu_ap}
f(u)&=&u^{4-2D}\left\{ u \int^{+\infty}_0 dt t^{\frac{D}{2}-1} 
K_{\frac{1-D}{2}}(t) \cos(\frac{t^{1/2}}{u}) \right.
\nonumber
\\
&+&\left. \frac{1}{2}
\int^{+\infty}_0 dt t^{\frac{D-3}{2}} K_{\frac{3-D}{2}}(t) 
\sin(\frac{t^{1/2}}{u}) \right\} \ ,
\eea
where $K_{\nu}$ is a modified Bessel function. 
We have not been able to compute the integrals in Eq.~\ref{def_fu_ap} 
explicitly, except in the case $D=2$. Nevertheless, we know both that $f(u)$ is
a monotonically increasing function of $u$ and its asymptotic behavior for
small and large $u$. For large $u$ we have the result,
\be\label{fu_largeu}
f(u)_{u\rightarrow \infty}=2^{\frac{D}{2}-2}\Gamma(\frac{1}{4})
\Gamma(\frac{D}{2}-\frac{1}{4})u^{5-2D}(1+{\cal O}(1/u))
\ee
while, for small $u$, we have
\bea\label{fu_smallu}
\lim_{u \rightarrow 0} f(u)&=&\frac{\pi \Gamma(\frac{3-D}{2})}
{2^{\frac{D+1}{2}}} \frac{1}{\Gamma(5-2D)\sin(\frac{\pi}{2}(5-2D))}
\nonumber\\
\lim_{u \rightarrow 0} f^{(n)}(u)&=&0 \ , \ n > 0
\eea
where $n$ stands for any derivative of $u$. The latter leads us to conjecture
that the corrections to Eq.~\ref{fu_smallu} are of the type
${\cal O}(e^{-\frac{1}{4u^2}})$, as explicitly seen at $D=2$.
A plot of $f(u)$ for different values of $D$ is given in Fig.~\ref{fig__fu}.

The next step is to compute $F(z)$ Eq.~\ref{def_FZ}. Its exact analytical 
form seems hopeless to compute, but again we can find its
asymptotic limits. For small $z$ we have
\be\label{FZ_smallz}
F(z)_{z \rightarrow 0}= \frac{\Gamma(\frac{2D-2}{5-2D})}
{(2^{D/2-2}\Gamma(1/4)\Gamma(D/2-1/4))^{\frac{2D-2}{5-2D}}}
\frac{z^{\frac{D-2}{5-2D}}}{5-2D}(1+{\cal O}(z^{1/(5-2D)}) \ ,
\ee
and for large $z$
\be\label{FZ_largez}
F(z) \sim e^{-f(0)z} \ ,
\ee
where $f(0)$ is given by Eq.~\ref{fu_smallu}.

\begin{table}[hbt]
\centerline{
\begin{tabular}{||r|l||r|l|}
\multicolumn{1}{c}{D} & \multicolumn{1}{c}{I(D)} &
\multicolumn{1}{c}{D} & \multicolumn{1}{c}{I(D)}     \\\hline
1.6 & $0.07951$         & 2.0 & $1.2639 10^{-3}$      \\\hline
1.7 & $0.04643$         & 2.1 & $1.0775 10^{-4}$      \\\hline
1.8 & $0.02027$         & 2.2 & $1.71557 10^{-6}$     \\\hline
1.9 & $0.006434$        & 2.3 & $4.01298 10^{-10}$    \\\hline
\end{tabular}}
\caption{Sample of values for $I(D)$.}
\label{tab__ID}
\end{table}

The asymptotics provide valuable cross checks for the  numerical
integrations, and we have also used them in speeding up
the numerical integration algorithms. A sample of values for $I(D)$ is
provided in Table \ref{tab__ID}.

\end{document}